\documentclass[12pt]{article}

\usepackage{a4wide}
\usepackage{amsmath}
\usepackage{bm}
\usepackage{amssymb}
\usepackage{epsfig}
\usepackage{epic}
\usepackage{mathrsfs}

\newcommand{\ft}[2]{{\textstyle\frac{#1}{#2}}}

\newcommand{\sfrac}[2]{{\textstyle\frac{#1}{#2}}}

\newcommand\re[1]{(\ref{#1})}
\def \res{\mathop{\rm res}\nolimits}

\def \baa {\begin{eqnarray*}}
\def \eaa {\end{eqnarray*}}

\newcommand{\be}{\begin{equation}}
\newcommand{\ee}{\end{equation}}
\newcommand{\ba}{\begin{eqnarray}}
\newcommand{\ea}{\end{eqnarray}}

\makeatletter
\def\mr@ignsp#1 {\ifx\:#1\@empty\else #1\expandafter\mr@ignsp\fi}%
\newcommand{\multiref}[1]{\begingroup%\let\protect\string%
\xdef\mr@no@sparg{\expandafter\mr@ignsp#1 \: }%
\def\mr@comma{}%
\@for\mr@refs:=\mr@no@sparg\do{\mr@comma\def\mr@comma{,}\ref{\mr@refs}}%
\endgroup}
\makeatother
\long\def\symbolfootnote[#1]#2{\begingroup%
\def\thefootnote{\fnsymbol{footnote}}\footnote[#1]{#2}\endgroup}

\newcommand{\Appref}[1]{Appendix~\multiref{#1}}

\numberwithin{equation}{section}
\begin{document}

%%%%%%%%%%%%%%%%%%%%%%%%%%%%%%%%%%%%%%%%%%%%%%%%%%%%%%%%%%%%%%%%%%%%%%
\thispagestyle{empty}

\begin{flushright}\footnotesize
\texttt{AEI-2009-066}\\
\vspace{0.5cm}
\end{flushright}
\setcounter{footnote}{0}

\begin{center}
{\Large\textbf{\mathversion{bold} Analytic solution of the multiloop Baxter
equation}}
\vspace{15mm}

{\sc M.~Beccaria$^a$, A.V.~Belitsky$^b$, A.V.~Kotikov$^c$  and
S.~Zieme$^{d,e}$}\\[10mm]

{\it $^a$ Physics Department, Salento University and INFN,\\
      73100 Lecce, Italy }\\[8mm]

{\it $^b$ Department of Physics, Arizona State University\\
          Tempe, AZ 85287-1504, USA}\\[8mm]

{\it $^c$ Bogoliubov Laboratory of Theoretical Physics\\
    Joint Institute for Nuclear Research\\
    141980 Dubna, Russia}\\[8mm]

{\it $^d$ Max-Planck-Institut f\"ur Gravitationsphysik\\
    Albert-Einstein-Institut \\
    Am M\"uhlenberg 1, D-14476 Potsdam, Germany}\\[8mm]

{\it $^e$ Faculty of Science, University of Herat, Afghanistan}\\[12mm]

\textbf{Abstract}\\[2mm]
\end{center}

\noindent{The spectrum of anomalous dimensions of gauge-invariant operators in maximally
supersymmetric Yang-Mills theory is believed to be described by a long-range integrable spin
chain model. We focus in this study on its $sl(2)$ subsector spanned by the twist-two
single-trace Wilson operators, which are shared by all gauge theories, supersymmetric or not.
We develop a formalism for the solution of the perturbative multiloop Baxter equation encoding
their anomalous dimensions, using Wilson polynomials as basis functions and Mellin transform
technique. These considerations yield compact results which allow analytical calculations of
multiloop anomalous dimensions bypassing the use of the principle of maximal transcendentality.
As an application of our method we analytically confirm the known four-loop result. We also
determine the dressing part of the five-loop anomalous dimensions.}

%%%%%%%%%%%%%%%%%%%%%%%%%%%%%%%%%%%%%%%%%%%%%%%%%%%%%%%%%%%%%%%%%%%%%%
\newpage

\allowdisplaybreaks

\setcounter{page}{1}
%%%%%%%%%%%%%%%%%%%%%%%%%%%%%%%%%%%%%%%%%%%%%%%%%%%%%%%%%%%%%%%%%%%%%%
\section{Introduction}
%%%%%%%%%%%%%%%%%%%%%%%%%%%%%%%%%%%%%%%%%%%%%%%%%%%%%%%%%%%%%%%%%%%%%%

The success of gauge theories in accurately describing the laws of nature is
based on the
availability of computational techniques, see e.g., Ref.\ \cite{VerMoc00}, which
allow for
a systematic improvement of approximations involved. Perturbative expansions in
the gauge
coupling constant $g_{\rm\scriptscriptstyle YM}$ are conventionally deduced from
Feynman
diagrams. However, due to uncontrollable proliferation of the latter at higher
orders
in $g_{\rm\scriptscriptstyle YM}$, the rules quickly become unmanageable, making
direct
computations already at four-loop order highly nontrivial and require massive
computer manipulations.
On top of this, individual Feynman diagrams obscure underlying properties of the
theory and reveal
simple results enjoying sometimes enhanced symmetries only in their sum. One was
therefore compelled
to search for an alternative approach which presented itself recently.

On the one hand, some time ago it was established that at weak coupling one-loop
spectra of anomalous
dimensions of maximal-helicity gauge-invariant operators in QCD coincide with
energy spectra of a
one-dimensional non-compact Heisenberg magnet \cite{BraDerMan98,Bel99}. The
latter can be diagonalized
by means of the traditional Bethe ansatz formalism of integrable systems and
yields anomalous dimensions
of the corresponding four-dimensional gauge theory. These simplifications are
echoed by higher loop
contributions, especially in supersymmetric gauge theories. It was found in
Refs.\ \cite{Lip98,BeiSta05}
that all single-trace operators in planar, maximally supersymmetric gauge theory
\be
\mathcal{O} = {\rm tr} \, \left( X (D_+^2 X) Y Z X \lambda X F_{+\perp} (D_+ Y)
\bar\lambda \dots \right)
\ee
can be described by a long-range integrable spin-chain model with elementary
excitations identified with
the particle fields $Y$, $Z$, $\lambda$ etc. of the gauge theory and/or
covariant derivatives $D_\mu$
acting on them propagating on the vacuum state $| 0 \rangle = {\rm tr} \, \left(
X^L \right)$. Less
supersymmetric Yang-Mills theories entertain integrability only in certain
closed subsectors under
renormalization group evolution \cite{BelDerKorMan04}.

On the other hand, the AdS/CFT correspondence \cite{Mal97} conjectures that the
strongly coupled
$\mathcal{N} = 4$ Yang-Mills theory is dual to a free type IIB super-string
theory on an AdS$_5
\times$S$^5$ background. The latter was found to be classically integrable as
well \cite{BenPolRoi03}.
Using this conjecture as a virtue led to a suggestion of an integrable
structure which interpolates
between weak and strong coupling regimes. Though the underlying spin chain model
is not known, a set of
Bethe ansatz equations is nevertheless available \cite{BeiSta05,BeiEdeSta06},
which has passed a number
of non-trivial tests at weak coupling, see e.g.~\cite{BeiEdeSta06} and
\cite{BerCzaDixKosSmi06}, as well as at strong coupling by positive comparison
 with perturbative string theory, see e.g.~\cite{BasKorKot07} and
\cite{BecForTirTse08}.

These findings suggest to use the putative integrable structure as an
alternative to the conventional
Feynman diagrams technique for multiloop calculations of anomalous dimensions.
In this paper, we develop
a practitioner's formalism building up on earlier considerations based on the
all-order Baxter equation
\cite{Bel06,Bel09} for finding the spectrum of twist-two Wilson operators
\be
\label{TwistTwo}
\mathcal{O} = {\rm tr} \, \left( X D_+^M X \right)
\, .
\ee
These arise in all gauge theories albeit with a different field content, the
scalar $X$ being specific
to supersymmetric cousins of QCD. Their anomalous dimensions have been obtained
diagrammatically to a
considerably high-order
\cite{KotLipVel03,MocVerVog04,KotLipOniVel04,FiaSanSieZan07,Vel08}.

The Baxter equation is advantageous over the Bethe ansatz formalism if one is
interested in a
systematic analytical framework. However they both enter on equal footings for
numerical studies,
and Bethe equations were used in the past together with the principle of maximal
transcendentality \cite{Kotikov:2002ab}
%\cite{KotLipOniVel04} 
to perform phenomenal computations
\cite{BajJanLuk08,BecForLukZie09}.

Our following consideration is a generalization of the study in Ref.\
\cite{KotRejZie08} which
was based on a deformation of the solution to the one-loop Baxter equation. What
will differ in
the current work is that we will introduce a new basis of functions used in the
construction of
next-to-leading order solutions, the so-called Wilson polynomials. For
comparison we also present the
basis of continuous Hahn polynomials used in \cite{KotRejZie08}. Furthermore we
obtain a new form for
non-polynomial contributions which is free from multiple sums involving Stirling
numbers. The
latter property is essential for obtaining analytical results for anomalous
dimensions in terms
of nested harmonic sums. Our subsequent presentation is organized as follows. In
the next section,
we briefly review the formalism of the Baxter equation in maximally
supersymmetric gauge theory
and then present a novel form of the solution in two- to four-loop order in the
gauge coupling.
The non-polynomial parts of the Baxter equation are analyzed in Section
\ref{non-polynomial} using
the Mellin transform technique. We present the analytic form of anomalous
dimensions and then in
Section \ref{five-loop dresci}, we discuss the reciprocity properties including
the dressing part
of the five-loop anomalous dimensions. Finally, we conclude. Several appendices
summarize basic
definitions required in the main body of the paper and details of calculations
which are two
lengthy to be presented in the main text.

%%%%%%%%%%%%%%%%%%%%%%%%%%%%%%%%%%%%%%%%%%%%%%%%%%%%%%%%%%%%%%%%%%%%%
\section{Baxter equation}
\label{Baxter equation}
%%%%%%%%%%%%%%%%%%%%%%%%%%%%%%%%%%%%%%%%%%%%%%%%%%%%%%%%%%%%%%%%%%%%%

The spin-chain description allows one to calculate anomalous dimensions of
Wilson operators as a
function of the 't Hooft coupling constant\footnote{Please note that we use
a different convention for the coupling constant then in \cite{KotRejZie08}.}
 $g^2 = g^2_{\rm\scriptscriptstyle YM}N_c/(4 \pi^2)$.
However, the formalism based on the Bethe ansatz equations has the drawback that
its predictions fail
when the order of the perturbative expansion in $g^2$ exceeds the length $L$ of
the operator under
study \cite{AmbJanKri05,KotLipRejStaVel07}. This implies that for twist-two
operators (\ref{TwistTwo})
the onset of wrapping effects occurs starting from four-loops already and the
complete anomalous
dimension is a sum of two terms
\be
\gamma (g) = \gamma^{(\rm asy)} (g) + \gamma^{(\rm wrap)} (g)
\, .
\ee
The first contribution $\gamma^{(\rm asy)}$ on the right-hand side is determined
by the solution to
the asymptotic Bethe ansatz equations and can be written in terms of the Baxter
function $Q(u)$ as
\cite{Bel06}
\be
\label{AllOrderAD}
\gamma^{(\rm asy)} (g) = i g^2 \int_{- 1}^1 \frac{dt}{\pi} \sqrt{1 - t^2}
\left(
\ln \frac{
Q \left( + \ft{i}{2} - g t \right)
}{
Q \left( - \ft{i}{2} - g t \right)
}
\right)^\prime
\, .
\ee
The latter is a degree-$M$ polynomial in the spectral parameter $u$ with zeros
determined by the
Bethe roots $u_k$
\be
\label{BaxterPolynomial}
Q (u) = \prod_{k = 1}^M (u - u_k(g))
\, .
\ee
It obeys an equation known as the asymptotic Baxter equation
\be
\label{BaxterEquation}
(x^+)^L {\rm e}^{\sigma_+ (u^+) - \Theta (u^+)} Q (u + i)
+
(x^-)^L {\rm e}^{\sigma_- (u^-) - \Theta (u^-)} Q (u - i)
=
t (u) Q (u)
\, .
\ee
Compared to the Baxter equation for the familiar non-compact nearest-neighbor
XXX Heisenberg
spin chain
\be
\label{1loopB}
\mathfrak{B} [Q_0]
\equiv
(u^+)^L Q_0 (u + i) + (u^-)^L Q_0 (u - i) - t_0 (u) Q_0 (u)
= 0
\, ,
\ee
with factors $(u^\pm)^L \equiv ( u \pm \ft{i}{2} )^L$ accompanying the
corresponding
Baxter polynomials, Eq.\ (\ref{BaxterEquation}) possesses highly non-trivial
dressing\footnote{This nomenclature should not be confused with the dressing
factor related to the phase $\Theta(u)$, which we also refer to in later sections.}
factors
reflecting coupling-constant dependent dynamics of the four-dimensional Yang-
Mills theory.
First, the spectral parameter gets renormalized \cite{BeiDipSta04} and reads $x
= x [u] =
\ft12 (u + \sqrt{u^2 - g^2})$, with the assumed conventional notation $x^\pm = x
[u^\pm]$,
and, second, the exponents $\sigma$ and $\Theta$ provide the interpolation
between weak and
strong-coupling expansions \cite{BeiEdeSta06} and read \cite{Bel09},
\ba
\label{Sigmap}
\sigma_\pm (u)
&=&
\int_{-1}^1
\frac{d t}{\pi}
\frac{\ln Q (\pm \ft{i}{2} - g t)}{\sqrt{1 - t^2}}
\left( 1 - \frac{\sqrt{u^2 - g^2}}{u + g t} \right) \, , \\
\Theta (u)
&=&
- 8 i \sum_{r=2}^\infty \sum_{s = r+1}^\infty
\left( \frac{g}{2} \right)^{r + s - 2} C_{rs} (g)
\int_{-1}^1 \frac{dt}{\pi} \sqrt{1 - t^2}
\left( \ln \frac{Q (+ \ft{i}{2} - g t)}{Q (- \ft{i}{2} - g t)} \right)^\prime
\nonumber\\
&&\qquad\qquad\qquad\qquad\qquad\times
\left\{
\left(- \frac{2}{g} \right)^{s - 2} \frac{U_{s - 2} (t)}{x^{r - 1}}
-
\left(- \frac{2}{g} \right)^{r - 2} \frac{U_{r - 2} (t)}{x^{s - 1}}
\right\}
\, ,
\ea
with the expansion coefficients given by
\be
C_{rs} (g) = \sin(\ft{\pi}{2} (s - r))
\int_0^\infty d v \frac{J_{r - 1} (gv) J_{s - 1} (gv)}{v ({\rm e}^v - 1)}
\, .
\ee
Since the Bethe roots acquire dependence on the 't Hooft coupling, the Baxter
function can be
expanded in a perturbative series $Q (u) = Q_0 (u) + g^2 Q_1 (u) + \dots$ and
each term found
explicitly as a solution to Eq.\ (\ref{BaxterEquation}) as we demonstrate next.
Notice that all
subleading Baxter functions $Q_{\ell > 0} (u)$ are polynomials in the spectral
parameter of a
degree two units lower than the leading $Q_0 (u)$. To four-loop order, the
dressing functions
admit the expansion
\ba
\label{ExpansionDressing}
\sigma^\pm (u )
&=&
g^2 \frac{i}{u} \gamma_0^\pm
+
g^4
\left[
\frac{i}{u} \gamma_1^\pm
-
\frac{1}{4 u^2} \left( (\gamma_0^\pm )^2 + \alpha^+ \right)
+
\frac{i}{4 u^3} \gamma_0^\pm
\right]
\\
&+& g^6
\bigg[
\frac{i}{u} \gamma_2^\pm
-
\frac{1}{u^2}
\left(
\ft{1}{2} \gamma_1^\pm \gamma_0^\pm + 2 \gamma_0^\pm \beta^\pm
- \ft{1}{4} (\alpha^\pm)^2 - \ft{1}{4} (\gamma_0^\pm)^2 \alpha^\pm + \chi^\pm
\right)
\nonumber\\
&&\ \
+
\frac{i}{u^3}
\left(
\ft{1}{4} \gamma_1^\pm
- \ft1{12} \gamma_0^\pm \left( (\gamma_0^\pm)^2 + \ft32 \alpha^\pm \right) +
\beta^\pm
\right)
-
\frac{1}{8 u^4} \left( (\gamma_0^\pm)^2 + \alpha^\pm \right)
+
\frac{i}{8 u^5} \gamma_0^\pm
\bigg]
\nonumber\\
&+& \dots
\, , \nonumber\\
\Theta (u) &=&
g^6 \, \zeta_3
\, \Re{\rm e}
\left[
\frac{1}{2 u^2} \gamma_0^+
-
\frac{i}{u} \left( (\gamma_0^+)^2 + \alpha^+\right)
\right]
+ \dots \, ,
\ea
where the expansion coefficients are introduced explicitly in Appendix
\ref{App:Inhomogeneities}.

Yet another unknown in Eq.\ (\ref{BaxterEquation}) is the transfer matrix, which
takes the form
\be
\label{TransferMatrix}
t (u)
=
\Re{\rm e} ( x^+ )^L
\left( 2 + \sum_{k \geq 1} \mathfrak{Q}_k (g) \Re{\rm e} ( x^+ )^{- k} \right)
-
\sum_{k \geq 1} \mathfrak{R}_k (g) \Im{\rm m} ( x^+ )^{- k}
\, .
\ee
Here the upper limits in the sums can exceed the length of the operator in
question and the emerging
charges $\mathfrak{Q}_{k > L}$ along with $\mathfrak{R}_k$ serve to compensate
non-polynomial terms
arising in the left-hand side of the finite difference equation
\re{BaxterEquation} stemming
from the expansion of the renormalized rapidity parameter and dressing factors
in Taylor series
in the 't Hooft coupling. The charges admit perturbative expansions
\ba
\mathfrak{Q}_k
&=& \mathfrak{Q}_k^{[0]} + g^2 \mathfrak{Q}_k^{[1]} + g^4 \mathfrak{Q}_k^{[2]}
+ \dots
\, , \\
\mathfrak{R}_k
&=& \mathfrak{R}_k^{[0]} + g^2 \mathfrak{R}_k^{[1]} + g^4 \mathfrak{R}_k^{[2]}
+ \dots
\, . \nonumber
\ea
And the only non-trivial contributions for $L = 2$ operators up to four-loop
order read
\ba
\mathfrak{Q }_2^{[0]}
&=& - M (M + 1) \, , \\
\mathfrak{Q}_2^{[1]}
&=& - (2 M + 1) \Re{\rm e} \left[ \gamma^+_0 \right]
\, , \nonumber\\
\mathfrak{Q}_2^{[2]}
&=& - (2 M + 1) \Re{\rm e} \left[ \gamma^+_1 \right]
- \ft12 \Re{\rm e} \left[ 3 (\gamma^+_0)^2 + \alpha^+ \right]
\, , \nonumber\\
\mathfrak{Q}_2^{[3]}
&=& - (2 M + 1) \Re{\rm e} \left[ \gamma^+_2 \right]
+ \ft12 \Re{\rm e} \left[ \alpha^+ \left( (\gamma^+_0)^2 + \alpha^+ \right)
- \gamma_0^+ \left( 6 \gamma_1^+ + 8 \beta^+ + \zeta_3 \right) - 4 \chi^+
\right]
\, , \nonumber\\
\mathfrak{R}_1^{[3]}
&=&
- \ft12 \Re{\rm e} \left[ \gamma^+_0 \left( (\gamma^+_0)^2 + \alpha^+ \right)
\right]
\, , \nonumber\\
\mathfrak{R}_2^{[3]} &=&
0
\, , \nonumber\\
\mathfrak{R}_3^{[3]}
&=&
\ft1{8} \Re{\rm e} \left[ \gamma^+_0 \right]
\, . \nonumber
\ea
Here $\mathfrak{Q}_2^{[0]}$ is the eigenvalue of the quadratic Casimir operator
of the collinear
conformal subgroup in the basis of conformal Wilson operators such that the
leading order transfer
matrix admits the conventional form for the two site non-compact Heisenberg spin
chain,
\be
t_0 (u) = (u^+)^2 + (u^-)^2 + \mathfrak{Q}_2^{[0]}
\, .
\ee
Finally, the solution to the Baxter equation has to be supplemented with the
condition of the
vanishing quasi-momentum
\be
i \vartheta
=
\frac{1}{\pi}
\int_{- 1}^1 \frac{dt}{\sqrt{1 - t^2}}
\ln \frac{Q (+ \ft{i}{2} - g t)}{Q (- \ft{i}{2} - g t)}
= 0
\, ,
\ee
in order to pick out only cyclic, physical states.

%%%%%%%%%%%%%%%%%%%%%%%%%%%%%%%%%%%%%%%%%%%%%%%%%%%%%%%%%%%%%%%%%%%%%%
\section{Wilson vs. Hahn}
%%%%%%%%%%%%%%%%%%%%%%%%%%%%%%%%%%%%%%%%%%%%%%%%%%%%%%%%%%%%%%%%%%%%%%

It is known for quite some time that the leading order solution $Q_0$ for the
non-compact
two-site Heisenberg magnet is given by the continuous Hahn
polynomials\footnote{We summarize
their basic properties in Appendix \ref{WilsonPolynomials}.} \cite{FadKor95},
\begin{equation}
\label{Q03F2}
Q_0(u)={}_3 F_2\left(\left. \begin{array}{c}
-M, \ M+1,\ \frac{1}{2}+iu \\
1,\  1 \end{array}
\right| 1\right) \,.
\end{equation}
Proceeding to higher loops, it was demonstrated in Ref.~\cite{KotRejZie08} that
subleading
contributions $Q_{\ell > 0} (u)$ to the Baxter function can be obtained by a
deformation of the
leading order result (\ref{Q03F2}). The equations which these corrections obey
remain of second order
in finite differences, but acquire inhomogeneous terms depending on lower-order
functions. This
implies that the structure of all polynomial higher-loop contributions can be
immediately understood
once the building blocks for the two-loop Baxter function are known. To find the
latter it suffices
to expand both side of the Baxter equation (\ref{BaxterEquation}) to
$\mathcal{O} (g^2)$ and find
\begin{eqnarray}
\left[ (u^+)^2 - \ft12 g^2 ( 1 - i \gamma_0 u^+ ) \right] Q (u + i)
&+&
\left[ (u^-)^2 - \ft12 g^2 ( 1 + i \gamma_0 u^- ) \right] Q (u - i)
\\
&=&
\left[ (u^+)^2 + (u^-)^2 - g^2 + \mathfrak{Q}_2 \right] Q(u)
\, , \nonumber
\end{eqnarray}
with the quadratic conformal Casimir $\mathfrak{Q}_2 \simeq - (M + 1 + \ft12 g^2
\gamma_0)(M + \ft12
g^2 \gamma_0)$ renormalized by the one-loop anomalous dimension $\gamma_0 = 2
\,S_1(M)$ to this
order of perturbation theory. Matching this to the equation obeyed by the
continuous Hahn
polynomials, we obtain the result for the Baxter function with incorporated 
two-loop corrections
\be
Q (u)
=
N (g)
{}_3 F_2
\left(
\left.
\begin{array}{c}
- M, \ M + 1 + g^2 \gamma_0 ,\ \frac{1}{2} + iu + \ft{i}{\sqrt{2}} g + \ft14 g^2
\gamma_0
\\[2mm]
1 + i \sqrt{2} g + \ft12 g^2 \gamma_0 , \ 1 + \ft12 g^2 \gamma_0
\end{array}
\right| 1 \right)
\, .
\ee
Expanding this in a Taylor series with respect to the 't Hooft coupling, we find
for the two-loop
correction itself
\be
\label{TwoLoopHahn}
Q_1 (u)
=
b_1 Q_0 (u)
+
\tfrac{1}{4}
\left(
2 \gamma_0 \partial_{\delta_1}
-
\partial_{\delta_2}^2
-
\partial_{\delta_3}^2
\right)
\left.
{}_3 F_2
\left(
\left.
\begin{array}{c}
- M, \ M + 1 + 2\,\delta_1 ,\ \frac{1}{2} + iu + \delta_2
\\[2mm]
1 + \delta_1 + \delta_2 + \delta_3, \ 1 + \delta_2 - \delta_3
\end{array}
\right| 1 \right)
\right|_{\delta_i = 0}
 ,
\ee
where we chose a spin-dependent form of the perturbative expansion of the
normalization constant
$N (g) \simeq 1 + g^2 b_1 (M)$ with
\begin{equation}
\label{NormQ1Hahn}
b_1 (M) = 4 S_1^2 + S_2 - 2 S_1\,\widetilde{S}_1 \, ,
\end{equation}
in order to reduce the degree of $Q_1$ in accordance with the definition
(\ref{BaxterPolynomial})
such that $\deg Q_1 = (\deg Q_0 - 2)$. Here and below the nested harmonic sums
(\ref{NestHarmSum})
appear as functions of two arguments
\baa
S_{a_1, a_2, \dots} \equiv S_{a_1, a_2, \dots}(M)
\, , \qquad
\widetilde{S}_{a_1, a_2, \dots}\equiv S_{a_1, a_2, \dots}(2M)
\, .
\eaa

The Baxter function for twist-two operators in the basis of continuous Hahn
polynomials has been
obtained analytically to three-loop order in \cite{KotRejZie08}. However, as can
be seen from Eq.\
(\ref{TwoLoopHahn}) the number of deformation terms arising is quite
substantial. This calls for
a quest to find a more concise representation. To this end one notices that the
Baxter function
for the ground state is symmetric under the reflection $u \to -u$, which is
however not transparent
in the representation in terms of Hahn polynomials (\ref{Q03F2}) but becomes
explicit in the basis
of the Wilson polynomials\footnote{We briefly review them in Appendix
\ref{WilsonPolynomials}.}
\begin{equation}\label{Q0}
Q_0 (u) ={}_4 F_3\left(\left.
    \begin{array}{c}
    -\frac{M}{2}, \ \frac{M+1}{2},\ \frac{1}{2}+iu,\ \frac{1}{2}-iu \\
    1,\ 1,\ \frac{1}{2}
    \end{array}
    \right| 1\right)\,.
\end{equation}
Analogously to the previous consideration, matching the two-loop Baxter equation
to the equation
for Wilson polynomials (\ref{WilsonEq}), we find the two-loop solution
\begin{equation}
Q (u)
= N (g)
{}_4 F_3
\left(
\left.
\begin{array}{c}
- \frac{M}{2}, \ \frac{M + 1}{2} + \ft12 g^2 \gamma_0 ,\ \frac{1}{2} + iu,\
\frac{1}{2} - iu
\\[2mm]
1 + \ft{i}{\sqrt{2}} g + \ft14 g^2 \gamma_0 , \ 1 - \ft{i}{\sqrt{2}} g + \ft14
g^2 \gamma_0 , \ \ft12
\end{array}
\right| 1 \right)
\, ,
\end{equation}
up to an overall coupling-dependent normalization constant $N (g)$. Expanding
this result to order
$g^2$ yields the two-loop Baxter polynomial $Q_1 (u)$ in Wilson basis
\begin{equation}
\label{Q1}
Q_1 (u)
=
a_1(M) Q_0(u)
+\ft14
\big( 2 \gamma_0 \partial_{\delta_1}
-
 \partial^2_{\delta_2} \big)
\,
{}_4 F_3 \left. \left(\left.
\begin{array}{c}
- \frac{M}{2}, \ \frac{M + 1}{2} + \delta_1 ,\ \frac{1}{2}+iu,\ \frac{1}{2}-iu
\\[1mm]
1 + \delta_1 + \delta_2 ,\ 1 - \delta_2 ,\ \frac{1}{2}
\end{array}
\right| 1\right) \right|_{\delta_1 , \delta_2 = 0}
\, ,
\end{equation}
where the normalization constant is $N (g) \simeq 1 + g^2 a_1 (M)$ with
\begin{equation}
a_1(M) = 3S_1^2 + S_2 + S_{-2} - 2S_1\,\widetilde{S}_1
\, .
\end{equation}
A few comments are in order concerning the relation of this representation to
the one in terms of
continuous Hahn polynomials. First of all, there is one deformation less. As a
consequence, the
number of polynomial contributions in higher loops will drastically decrease.
Second, the deformed
parts in the two representations are not identical and hence their degree
reduction coefficients
multiplying the leading order solution differ as well. However, it should be
noted that these degree reduction coefficients can, in both cases, not contribute
to the anomalous
dimension by symmetry arguments. Although less obvious, the same is true for
higher-loop contributions.

Let us now turn to higher loops contributions and use the finding of this
section to devise an
efficient formalism to determine perturbative solutions. We will present the
results both in the
basis of Wilson and continuous Hahn polynomials.

%%%%%%%%%%%%%%%%%%%%%%%%%%%%%%%%%%%%%%%%%%%%%%%%%%%%%%%%%%%%%%%
\section{Polynomial contributions}
\label{PolinomialTerms}
%%%%%%%%%%%%%%%%%%%%%%%%%%%%%%%%%%%%%%%%%%%%%%%%%%%%%%%%%%%%%%%

To start with, as we observed in the previous section in order to tackle higher
order
corrections to the Baxter function it suffices to introduce a doubly-deformed
function
and its derivatives with respect to the deformation parameters
\be
T_{(p, q)}
\equiv
\partial^p_{\delta_1} \partial^{2 q}_{\delta_2}
\left.
{}_4F_3
\left( \left.
\begin{array}{c}
- \frac{M}{2} \ , \ \frac{M+1}{2} + \delta_1 \ ,\ \frac{1}{2} + i\,u \ ,\
\frac{1}{2} - i\,u
\\[1mm]
1 + \delta_1 + \delta_2 \ ,\ 1 - \delta_2 \ ,\ \frac{1}{2}
\end{array}\right| 1\right)
\right|_{\delta_1 = \delta_2 = 0}
\, .
\ee
Then a straightforward scheme presents itself for the construction of the $\ell-
$th order function
$Q_\ell (u)$. That is, $Q_\ell (u)$ is a linear superposition of the structures
$T_{(\ell_1, \ell_2)}$ with $\ell_1
+ \ell_2 \leq \ell$ accompanied by degree-$d$ transcendental numbers of $d = 2
\ell - 2 \ell_2 -
\ell_1$ and a $Q_0$-proportional term, such that the degree of $Q_\ell$ is
reduced to
$\deg Q_\ell = (\deg Q_0 - 2)$.

Analogously, in the basis of continuous Hahn polynomials we have a triple-
deformed function
\be
\label{HahnDeformation}
T_{(p,q,r)} =
\partial_{\delta_1}^p \partial_{\delta_2}^{q} \partial_{\delta_3}^{r}
\left.
{}_3 F_2
\left(
\left.
\begin{array}{c}
- M, \ M + 1 + \delta_1 ,\ \frac{1}{2} + iu + \delta_2
\\[2mm]
1 + \delta_1 + \delta_2 + \delta_3, \ 1 + \delta_2 - \delta_3
\end{array}
\right| 1 \right)
\right|_{\delta_1 = \delta_2 = \delta_3 = 0}
\, .
\ee

In the basis of Hahn polynomials there are more possible deformations than in
the Wilson basis.
As a consequence the number of terms at each order of the perturbative series is
also considerably
increased. As was shown in Ref.\ \cite{KotRejZie08} at three loops, a simple
counting\footnote{Note,
that for this representation, there is a term $T_{(0,0,3)}$, which has a third-
order derivative
w.r.t.\ the deformation parameter.} gives a total number of eleven terms.
Changing the representation
from ${}_3F_2 \to {}_4F_3$ reduces the number of contributions to the Baxter
function by almost half.
At four-loop order this effect will decrease the number of contributions from
$36$ to $15$ terms, as
demonstrated below.

In the following we will divide the contributions to the Baxter functions into
polynomial and non-polynomial contributions
\begin{equation}\label{split of contributions}
Q_\ell (u) = Q_\ell^{(p)} (u) + Q_\ell^{(np)} (u)
\, ,
\end{equation}
respectively. The terminology used here needs clarification. Of course, at any
given order of
perturbation theory, the Baxter equation is polynomial. However, it consists of
two types of
terms, the first one explicitly polynomial in the spectral parameter $u$ and the
other containing
inverse powers of the spectral parameter $u^\pm$ accompanying Baxter polynomials
and thus appearing
superficially non-polynomial. Indeed the inverse powers of $u$ conspire to cancel
in the sum of
the latter such that the net result is polynomial as it should. However we
choose to split the
Baxter function according to this nomenclature inherited from their source in
the equation.

%%%%%%%%%%%%%%%%%%%%%%%%%%%%%%%%%%%%%%%%%%%%%%%%%%%%%%%%%%%%%%%
\subsection{Wilson basis}
%%%%%%%%%%%%%%%%%%%%%%%%%%%%%%%%%%%%%%%%%%%%%%%%%%%%%%%%%%%%%%%

Following the strategy outlined above we find first the polynomial part of the
perturbative
Baxter function in the Wilson basis. First, the three-loop Baxter function
$Q^{(p)}_2$ reads
\begin{eqnarray}
\label{Q2}
Q_2^{(p)} (u)
=
a_2 \, Q_0(u)
&+&
\ft12 \left( \gamma_1 + a_1 \gamma_0 \right) T_{(1,0)}
-
\ft18 \left( K_2 + 2 a_1 \right) T_{(0,1)}
\nonumber\\
&+& \ft18 \gamma_0^2 T_{(2,0)}
- \ft18 \gamma_0 T_{(1,1)}
+ \ft1{96} T_{(0,2)}
\, ,
\end{eqnarray}
where the transcendental coefficient $K_2$ of degree two is a linear superposition
of anomalous dimensions and inhomogeneities introduced in \Appref{App:Inhomogeneities}
\begin{equation}
\label{K_2}
K_2 = \alpha + \ft34 \gamma_0^2 = \ft12 \gamma_0^2 - S_{-2}
\, ,
\end{equation}
and the normalization function $a_2=a_2(M)$ reducing the degree of the higher-loop
polynomial depends on the non-polynomial contribution computed later in Section
\ref{non-polynomial}.

\medskip

At four-loop order a further transcendental function arises from the Baxter
equation, see
\Appref{App:Inhomogeneities}, from the expansion coefficients of the dressing
factors
\eqref{ExpansionDressing}. They appear in a certain combination with a degree of
transcendentality
four,
\begin{equation}
\label{K_4}
K_4
= -6\alpha^2+24\chi-\varepsilon\gamma_{0}+\delta\gamma_{0}^2
+
\tfrac{1}{4}\gamma_{0}^4+10\gamma_{0}\gamma_{1}\,.
\end{equation}
So that finally the polynomial part of the four-loop Baxter function is
\begin{eqnarray}
\label{Q3}
Q_3^{(p)} &(u)
&=
a_3 \, Q_0 (u)
+
\ft12
\left( \gamma_2 + a_1 \gamma_1 + a_2 \gamma_0 \right) T_{(1,0)}
-
\ft1{48}
\left( K_4 + 3 \zeta_3 \gamma_0 + 6 a_1 K_2 + 12 a_2 \right) T_{(0,1)}
\nonumber\\
&+&
\ft18 \left( 2 \gamma_1 \gamma_0 + a_1 \gamma_0^2 \right) T_{(2,0)}
-
\ft18
\left( \gamma_1 + a_1 \gamma_0 + \ft{1}{2} K_2 \gamma_0 \right) T_{(1,1)}
+
\ft1{192}
\left( \gamma_0^2 + 2K_2 + 2 a_1 \right) T_{(0,2)}\nonumber\\
&+&
\ft1{48} \gamma_0^3 T_{(3,0)}
-
\ft1{32} \gamma_0^2 T_{(2,1)}
+
\ft1{192} \gamma_0 T_{(1,2)}
-
\ft1{5760} T_{(0,3)}
\, ,
\end{eqnarray}
with the degree-reducing coefficient $a_3$. In the last equation, the term proportional to $\zeta_3$
stems from the dressing factor.

%%%%%%%%%%%%%%%%%%%%%%%%%%%%%%%%%%%%%%%%%%%%%%%%%%%%%%%%%%%%%%%
\subsection{Hahn basis}
%%%%%%%%%%%%%%%%%%%%%%%%%%%%%%%%%%%%%%%%%%%%%%%%%%%%%%%%%%%%%%%

Let us also include for completeness the three-loop Baxter function obtained in
\cite{KotRejZie08},
converted to the notation of Eq.\ \eqref{HahnDeformation} and rescaled coupling,
\begin{eqnarray}
\label{Q2Hahn}
Q_2^{(p)} (u)&=&b_2(M)Q_0(u)+\ft12 \left( \gamma_1 + b_1 \gamma_0 \right)
T_{(1,0,0)}
          -\ft18 \left(  K_2 + 2 b_1 \right)
(T_{(0,2,0)}+T_{(0,0,2)})\nonumber\\
&&+ \ft18 \gamma_0^2 T_{(2,0,0)}
- \ft18 \gamma_0 (T_{(1,2,0)}+T_{(1,0,2)}+\tfrac{1}{3}T_{(0,0,3)})\nonumber\\
&&+ \ft1{96}(6 T_{(0,2,2)}+ T_{(0,4,0)}+T_{(0,0,4)})
\, .
\end{eqnarray}
The appearing functions are the two-loop normalization constant for the Hahn
basis $b_1$ given
in \eqref{NormQ1Hahn} and the transcendental function $K_2$ in \eqref{K_2}. We
omit the precise
structure of the normalization constant $b_2(M)$, as it will be of no use for
us.

\medskip

In the basis of continuous Hahn polynomials the four-loop result is quite
lengthy and reads in the
conventions of Eq.~\eqref{HahnDeformation},
\begin{eqnarray}
\label{Q3Hahn}
Q_3^{(p)} &=& b_3 Q_0(u)+ \ft12 \left( \gamma_2 + b_1 \gamma_1 + b_2 \gamma_0
\right) T_{(1,0,0)}
\nonumber\\
&&- \ft1{48}\left(K_4 + 3 \zeta_3 \gamma_0 + 6 b_1 K_2 + 12 b_2\right)\left(
T_{(0,2,0)} + T_{(0,0,2)} \right)
\nonumber\\
&&+\ft1{192} K_2 \left( T_{(0,0,4)} + T_{(0,4,0)} + 6 T_{(0,2,2)} \right)
\nonumber\\
&&-\ft1{16} \gamma_0 \left( K_2 - 2 \gamma_1 + 2 b_1 \right)
\left( T_{(1,2,0)} + \ft13 T_{(0,3,0)} + T_{(1,0,2)} \right)
\nonumber\\
&&+
\ft18 \gamma_0 \left( \gamma_1 + b_1 \gamma_0 \right) T_{(2,0,0)}
+\ft1{192} \left( 2 b_1 + K_2 \right)\left( T_{(0,0,4)} + T_{(0,4,0)} + 6
T_{(0,2,2)} \right)
\nonumber\\
&&-
\ft{1}{5760}\left(T_{(0,0,6)} + T_{(0,6,0)} + 15 T_{(0,2,4)} + 15
T_{(0,4,2)}\right)
\nonumber\\
&&+\ft1{32} \gamma_0\left(T_{(1,2,2)} + \ft16 T_{(1,0,4)} + \ft13 T_{(0,3,2)}
+ \ft16 T_{(1,4,0)} + \ft1{15} T_{(0,5,0)}\right)
\nonumber\\
&&- \ft1{32} \gamma_0^2
\left(
T_{(2,0,2)} + T_{(2,2,0)} + \ft16 T_{(0,4,0)}
- \ft16 T_{(0,0,4)} + \ft23 T_{(1,3,0)}
\right)
+
\ft1{48} \gamma_0^3 T_{(3,0,0)}
\, .
\end{eqnarray}
Again, we omit the definition of the normalization $b_3$. $K_4$ is determined by
\eqref{K_4}. These
expressions for the three- and four-loop Baxter functions \eqref{Q2Hahn} and
\eqref{Q3Hahn} in the
Hahn basis should be compared to their Wilson basis counterparts in Eqs.\
\eqref{Q2} and \eqref{Q3},
respectively. It is apparent that the Wilson basis enormously simplifies higher-
loop computations
and should be the method of choice for subsequent loop-orders.

%%%%%%%%%%%%%%%%%%%%%%%%%%%%%%%%%%%%%%%%%%%%%%%%%%%%%%%%%%%%%%%
\section{Non-polynomial contributions}
\label{non-polynomial}
%%%%%%%%%%%%%%%%%%%%%%%%%%%%%%%%%%%%%%%%%%%%%%%%%%%%%%%%%%%%%%%

In order to complete the solution to the Baxter equation we have to address the
non-polynomial
contributions. According to the nomenclature of Section \ref{PolinomialTerms},
we collectively
label all inhomogeneities of the form $(u^\pm)^{-k} Q_\ell (u)$ with $k>1$ as
non-polynomial.
Recall however, that the transfer matrix (\ref{TransferMatrix}) is chosen in
such a way, that it
compensates these non-polynomialities and in the sum of all these contributions
at a given
loop-order the polynomiality is restored.

Non-polynomial inhomogeneities in the Baxter equation of twist-two operators
appear for the first
time at three-loop order, while wrapping effects set in at four loops. According
to our choice
of splitting in \eqref{split of contributions} we will complete the three-loop
Baxter function by
obtaining the term $Q_{\ell=2}^{(np)}$ in an novel form. The non-polynomial
contributions at three-loop
have already been found in \cite{KotRejZie08}. However, the representation used
there is given in
terms of Stirling numbers, which complicates the computation of anomalous
dimension. Therefore we
will present a novel representation of these terms here, which is based solely
on Mellin transform
techniques of the $(u^\pm)^{-k}$ with $k>1$. The procedure is different to the
one presented in
\cite{KotRejZie08}, as there the effective polynomial of all non-polynomial
terms has been expressed
in terms of Stirling numbers. Successively its contribution to the Baxter
function has been obtained
by Mellin transform\footnote{This procedure is given as Lemma 1 in
\cite{KotRejZie08}.}. What will
be different in our novel representation is that we give the Mellin transform
of all single
non-polynomialities. This results in a toolbox, which allows to simply construct
the solution from
a general set of expressions. The advantage is, that the final representation
does not depend on
complicated coefficient functions involving Stirling numbers, rendering the
computation of anomalous
dimension more feasible.

Before we turn to the explanation of the method used in our calculation, let us
summarize the
results of our analysis in this section. The non-polynomial contributions to the
three- and
four-loop Baxter functions are given by the following expressions
\begin{eqnarray}
\label{Q2np}
Q_2^{(np)} (u) &=&
\sum_{k = 0}^M \,2\,\Re{\rm e} \left[ P_k(u) \right] \, R_{k}(M) \,
r_{2,k}^{(np)} (M)
\, , \\
\label{Q3np}
Q_3^{(np)} (u) &=&
\sum_{k = 0}^M \,2\,\Re{\rm e} \left[ P_k(u) \right] \, R_{k}(M)
\left(
r_{3,k}^{(npp)} (M) + r_{3,k}^{(pnp)} (M)
\right)
\, ,
\end{eqnarray}
with expansion coefficients determined in Eqs.\ (\ref{r2k}), and (\ref{N4b9}),
(\ref{N4d7}), (\ref{N4c7})
respectively. These expressions complete the three- and four-loop Baxter
function, which is
given by the sum of \eqref{Q2} and \eqref{Q2np}, and (\ref{Q3}) and
(\ref{Q3np}), respectively.
The functions used in the representation of the non-polynomial parts are given
by
\begin{equation}
\label{N1}
R_k(M) \, = \, \frac{(- 1)^k}{(k !)^2}
\frac{\Gamma(M + 1 + k)}{\Gamma(M + 1 - k)}\,, \quad
P_k(u) \, = \,  \frac{\Gamma(k + \ft12 + iu)}{k!\Gamma(\ft12 + iu)}\,.
\end{equation}
Details of the computation are presented in the following subsections, while the
complete
dissection of the Mellin techniques is deferred to
\Appref{MellinTransformation}.

%%%%%%%%%%%%%%%%%%%%%%%%%%%%%%%%%%%%%%%%%%%%%%%%%%%%%%%%%%%%%%%
\subsection{Three loops}
\label{npThreeloop}
%%%%%%%%%%%%%%%%%%%%%%%%%%%%%%%%%%%%%%%%%%%%%%%%%%%%%%%%%%%%%%%

At three loops order, the non-polynomial contribution $U_2 (u)$ to the Baxter
equation
\be
\label{3loopB}
\mathfrak{B} [Q_2] = \dots + U_2 [Q_0]
\, ,
\ee
reads
\begin{equation}
\label{U2}
U_2 [Q_0]=
\frac{1}{16 (u^+)^2} \left[ Q_0 (u + i) - Q_0 (u) \right]  + \frac{i
\gamma_0^+}{4 u^+} Q_0 (u + i)
+ {\rm c.c.}\, .
\end{equation}
The procedure for finding a closed solution to this equation via the Mellin
transform technique
together with a complete set of building blocks required for generic higher
order analysis of
non-polynomial contributions is presented in \Appref{MellinTransformation}. We
refer to it for
a complete list of definitions of objects arising in this calculation. Below we
merely assemble
specific terms at three-loop order.

Extracting the first, constant term from $Q_0$ accompanying $\gamma_0$ in Eq.\
(\ref{U2}) and
introducing a new function $\widetilde Q_0$ via Eq.\ (\ref{M22}), we can
decompose $U_2$
into a sum of two terms with each of them being separately polynomial. Then,
making use of Eqs.\
(\ref{M38}), (\ref{M40}) and (\ref{M42}) we can obtain solutions to Eq.\
(\ref{3loopB}) stemming
separately from both combinations of non-polynomial contributions.

The first inhomogeneity in $U_2$ is of the form of (\ref{M38}) with $L=0$, i.e.,
\begin{equation}
\label{N2}
\frac{1}{u^+} \, \widetilde{Q}_0(u + i) -  \frac{1}{u^-} \, \widetilde{Q}_0(u-
i)
\,   = \, i
\sum_{p=1}^{M} \, \frac{2\,\Re{\rm e} [P_{p-1}(u)]}{p}  \, R_{p}(M) \,.
\end{equation}
Then, matching the right-hand side of this equation to Eq.\ (\ref{M18}) provides
via Eq.\
(\ref{M21}) the contribution of this inhomogeneous term to the Baxter function
$Q_2 (u)$,
which reads
\be
\label{N3}
i \sum_{k=0}^{M}  \,2\, \Re{\rm e}[P_{k}(u)] \, R_{k}(M) \, S_3(k)
\, ,
\ee
up to an overall factor $\ft{i}{8} \gamma_0 = \pm \ft{i}{4} \gamma_0^\pm$
accompanying (\ref{N2})
in Eq.\ (\ref{U2}).

Now turning to the second inhomogeneity in $U_2$, one notices that it is given
as a linear
combination of Eqs.~(\ref{M38}), (\ref{M38a}) and (\ref{M40}) with $L=1$
and $\widetilde{Q}_{0}(\pm \ft{i}{2})=0$
\begin{eqnarray}
\label{N3a}
\frac{1}{(u^+)^2} \, \widetilde{Q}_0(u+ i)
+
\frac{1}{(u^-)^2} \, \widetilde{Q}_0(u - i)
&=&
\, - i Q^\prime_{0}( \ft{i}{2}) \,
\sum_{p=M+1}^{\infty}  \frac{2\,\Re{\rm e}[P_{p-1}(u)]}{p}
\nonumber\\
&+&
\sum_{p=1}^{M} \frac{2 \,\Re{\rm e}[P_{p-1}(u)]}{p} \,
\sum_{k=1}^{p} \frac{R_{k}(M)}{k}
\, ,  \\
\label{N3a1}
\frac{1}{(u^+)^2} \, \widetilde{Q}_0(u)
+
\frac{1}{(u^-)^2} \, \widetilde{Q}_0(u)
&=&
\, - i Q^\prime_{0}(-\ft{i}{2}) \,
\sum_{p=M+1}^{\infty} \frac{2\,\Re{\rm e}[P_{p-1}(u)]}{p}
\nonumber\\
&-&
\sum_{p=1}^{M} \frac{2\,\Re{\rm e}[P_{p-1}(u)]}{p} \, \sum_{k=1}^{p} \,
R_{k}(M) \, Z_1(p,k)
\, ,
\end{eqnarray}
and (\ref{M42}) with $L=0$
\begin{eqnarray}
\label{N3b}
\frac{1}{u^+} - \frac{1}{u^-}
&=& - i \,
\sum_{p=1}^{\infty}  \frac{ 2\, \Re{\rm e}[P_{p-1}(u)]}{p}
\, ,
\end{eqnarray}
respectively. Then, the sought-after combination of terms is
\baa
\frac{1}{(u^+)^2}  \Big( \widetilde{Q}_0(u + i) &-& \widetilde{Q}_0(u) \Big)
+  \frac{1}{(u^-)^2}  \Big( \widetilde{Q}_0(u- i)- \widetilde{Q}_0(u) \Big)
+ 2 i \gamma_0 \left(\frac{1}{u^+}  -  \frac{1}{u^-}\right)
\\
&=&
\sum_{p=1}^{M} \, \frac{2 \, \Re{\rm e}[P_{p-1}(u)]}{p} \,
\sum_{k=1}^{p} \, R_{k}(M) \, Z_1(p,k-1)
\nonumber\\
&-& i \, \Big( Q^\prime_{0}(\ft{i}{2}) -  Q^\prime_{0}(-\ft{i}{2}) \Big)
\sum_{p=M+1}^{\infty}  \frac{2 \, \Re{\rm e}[P_{p-1}(u)]}{p}
+ 2 \gamma_0 \sum_{p=1}^{\infty}  \frac{2 \, \Re{\rm e}[P_{p-1}(u)]}{p}
\,  \nonumber
\eaa
And since $ i Q^\prime_{0}(\pm \ft{i}{2}) = \pm \gamma_0$, the contribution of
the infinite
series $\sim \sum_{p=M+1}^{\infty}$ cancels between the last two terms and the
result takes
a polynomial form
\ba
\label{N6}
\frac{1}{(u^+)^2}  \Big( \widetilde{Q}_0(u + i) &-& \widetilde{Q}_0(u) \Big)
+  \frac{1}{(u^-)^2}  \Big( \widetilde{Q}_0(u- i)- \widetilde{Q}_0(u) \Big)
+ 2 i \gamma_0 \left(\frac{1}{u^+}  -  \frac{1}{u^-}\right)
\\
&=&
\sum_{p=1}^{M} \, \frac{2 \, \Re{\rm e}[P_{p-1}(u)]}{p} \,
\biggl\{
\sum_{k=1}^{p} \, R_{k}(M) \, Z_1(p,k-1) + 2 \gamma_0
\biggr\}
\, . \nonumber
\ea
Combining this result with the one found earlier in Eq.\ (\ref{N3}) multiplied
by its proper
relative coefficient in Eq.\ (\ref{U2}) and identifying the expansion
coefficients in the
summand with $B_k$ of Eq.~(\ref{M20}), we can immediately write down the
contribution of both
inhomogeneous terms to $Q_2$ as
\ba
16  \, \sum_{k=0}^{M} 2 \, \Re{\rm e}[ P_{k}(u)] \, R_{k} (M) r_{2,k}^{(np)}(M)
\, \, ,
\label{N7}
\ea
with $r_{2,k}^{(np)}$, introduced in Eq.\ (\ref{M21}), taking the following
explicit form
\ba
\label{r2k}
r_{2,k}^{(np)}(M)
&=&
\ft{1}{16}
\Big(
\widehat{V}_{3,1}(k) + \widetilde{V}_{3,1}(k) + 2 \gamma_0 (W_3(k) - S_3(k))
\Big)
\nonumber \\
&=&
\ft{1}{16}
\Big(
V_{3,1}(k) + 2 \gamma_0 (W_3(k) - S_3(k))
\Big)
\, ,
\label{N7a}
\ea
determined in terms of the following sums
\ba\label{N8}
W_a(k)
&=&
\sum_{l=1}^{k} \, \frac{1}{l^a \, R_l(M)}
\, , \\
V_{a,b}(k)
&=&
\sum_{r=1}^{k} \, \frac{1}{r^a \, R_r(M)}
\, \sum_{m=1}^{r} \, R_m(M) \, Z_b(r,m-1)
\, ,  \\
\widehat{V}_{a,b}(k)
&=&
\sum_{r=1}^{k} \, \frac{1}{r^a \, R_r(M)}
\, \sum_{m=1}^{r} \, \frac{R_m(M)}{m} \, Z_{b-1}(r,m-1)
\, ,  \\
\label{N8last}
\widetilde{V}_{a,b}(k)
&=&
\sum_{r=1}^{k} \, \frac{1}{r^a \, R_r(M)} \, \sum_{m=1}^{r} \, R_m(M) \,
Z_b(r,m)
\, ,
\ea
with $Z_b$ introduced in Eq.\ (\ref{M36}). Restoring the overall coefficient
$\tfrac{1}{16}$
of (\ref{N6}) in Eq.\ (\ref{U2}), we recognize \eqref{N7} as Eq.\ \eqref{Q2np}.

%%%%%%%%%%%%%%%%%%%%%%%%%%%%%%%%%%%%%%%%%%%%%%%%%%%%%%%%%%%%%%%
\subsection{Four loops}
\label{npFourloop}
%%%%%%%%%%%%%%%%%%%%%%%%%%%%%%%%%%%%%%%%%%%%%%%%%%%%%%%%%%%%%%%

At four loops the number of non-polynomial contributions is greater, however,
due to the
perturbative iteration, the contribution of inhomogeneities due to the two-loop
Baxter
function $Q_1$ has the same form as in Eq.\ (\ref{U2}), i.e., $U_2 [Q_1]$. The
complete
set of non-polynomial inhomogeneities then consists of three terms,
\be
\mathfrak{B} [Q_3] = \dots + U_2 [Q_1] + U_3 [Q_0] + U_1 [Q_2^{(np)}] \, ,
\ee
where $U_1[Q_2^{(np)}]$ arises from non-polynomial contributions to the three-
loop Baxter function
$Q_2$ computed above and has the form
\ba\label{N4c1}
U_1 [Q_2^{(np)}]
&=& \left( \ft{1}{2} \, - i \gamma_0^+ u^+ \right) Q_2^{(np)} (u+i)
+ \left( \ft{1}{2} \, - i \gamma_0^- u^- \right) Q_2^{(np)} (u-i)
\nonumber \\
&& - \left(1+(2M+1)\gamma_0^+ \right) Q_2^{(np)} (u)
%{\rm c.c.}
\, .
\ea
While $U_3$ is another novel non-polynomial function of the leading order $Q_0$,
\ba
\label{U3}
U_3 [Q_0]
&=& \biggl\{
\frac{1}{32 (u^+)^4}
\Big( Q_0 (u + i) - Q_0 (u) \Big)
+ \frac{i\gamma_0^+ }{16 (u^+)^3}
\Big( Q_0 (u + i) + Q_0 (u) \Big)
\\
&+& \frac{iB_1^+}{24 u^+}
Q_0 (u + i)
-
\frac{i}{4 u^+} \gamma_0^+
\left( \alpha^+ + (\gamma_0^+)^2 \right) Q_0 (u)  \biggl\}
\,
+ \, {\rm c.c.}
\, , \nonumber
\ea
where
\be
B_1^{\pm} = 10 (\gamma_0^{\pm})^3 + 4 \gamma_1^{\pm} + 6 \gamma_0^{\pm}
\alpha^{\pm}
- 2  \gamma_0^{\pm} \delta^{\pm}  + \varepsilon^{\pm}
\, .
\ee
Let us address all three contributions in turn, starting with the latter.

%%%%%%%%%%%%%%%%%%%%%%%%%%%%%%%%%%%%%%%%%%%%%%%%%%%%%%%%%%%%%%%
\subsubsection{Inhomogeneity $U_3$}
%%%%%%%%%%%%%%%%%%%%%%%%%%%%%%%%%%%%%%%%%%%%%%%%%%%%%%%%%%%%%%%

Following the methodology developed at three-loop order, we split the
inhomogeneity $U_3$
into a sum of terms whose series representations can be matched into the generic
types
analyzed in Appendix \ref{MellinTransformation}. According to results given
there, the
inhomogeneity in Eq.\ (\ref{U3}) is a linear combination of Eqs.~(\ref{M38}),
(\ref{M40})
with $L=0, 2, 3$ and $\widetilde{Q}_{0}(\pm \ft{i}{2})=0$, namely,
\begin{eqnarray}
\label{N4a1}
\frac{1}{(u^+)^4} \, \Big( \widetilde{Q}_0(u+ i) &-& \widetilde{Q}_0(u) \Big)
+
\frac{1}{(u^-)^4} \,  \Big( \widetilde{Q}_0(u - i) - \widetilde{Q}_0(u) \Big)
\\
= &-& \sum_{p=1}^{M} \frac{2 \,\Re{\rm e}[P_{p-1}(u)]}{p} \,
\sum_{k=1}^{p} \, R_{k}(M) \Big( \tfrac{1}{2k} Z_2(p,k-1) + \tfrac{1}{6}
Z_3(p,k) \Big)
\nonumber \\
&+& \sum_{p=M+1}^{\infty} \frac{2\,\Re{\rm e}[P_{p-1}(u)]}{p}
\, \Big\{  \ft{i}2 \left(  Q^\prime_{0}(-\ft{i}{2}) -  Q^\prime_{0}(\ft{i}{2})
\right) Z_2(p)
\nonumber \\
&&\qquad
+
\tfrac{1}{2}
\left(
Q^{\prime\prime}_{0}(-\ft{i}{2}) -  Q^{\prime\prime}_{0}(\ft{i}{2})
\right) Z_1(p)
+
\ft{i}{6} \left(  Q^{\prime\prime\prime}_{0}(-\ft{i}{2})
-
Q^{\prime\prime\prime}_{0}(\ft{i}{2}) \right)
\Big\}
\, , \nonumber
\end{eqnarray}
multiplied by the factor of $\ft{1}{32}$. The term
\begin{eqnarray}
\label{N4a2}
&&
\frac{i}{(u^+)^3} \,
\Big( \widetilde{Q}_0(u + i) + \widetilde{Q}_0(u) \Big)
-
\frac{i}{(u^-)^3} \,
\Big( \widetilde{Q}_0(u - i) + \widetilde{Q}_0(u) \Big)
\\
&&\qquad\qquad\qquad\qquad
=
\sum_{p=1}^{M} \frac{2 \,\Re{\rm e}[P_{p-1}(u)]}{p} \,
\, \sum_{k=1}^{p}
\, R_{k}(M) \Big( \tfrac{1}{k} Z_1(p,k-1) - \tfrac{1}{2} Z_2(p,k) \Big)
\nonumber \\
&&\qquad
-
\sum_{p=M+1}^{\infty} \frac{2\,\Re{\rm e}[P_{p-1}(u)]}{p}
\, \Big\{  i \Big(  Q^\prime_{0}(-\ft{i}{2}) +  Q^\prime_{0}(\ft{i}{2}) \Big)
Z_1(p)
- \ft{1}{2} \Big(  Q^{\prime\prime}_{0}(-\ft{i}{2}) +
Q^{\prime\prime}_{0}(\ft{i}{2}) \Big)
\Big\}
\, , \nonumber
\end{eqnarray}
accompanied by $\ft{1}{16} \gamma_0^+$. And the terms
\ba
\label{N4a4}
&&
\frac{i}{u^+} \widetilde{Q}_0(u+ i) - \frac{i}{u^-} \widetilde{Q}_0(u- i)
\, = \,
-
\sum_{p=1}^{M} \frac{2\,\Re{\rm e}[P_{p-1}(u)]}{p} \, R_p(M) \, , \\
\label{N4a5}
&&
\frac{i}{u_+} \widetilde{Q}_0(u) - \frac{i}{u_-} \widetilde{Q}_0(u)
\, = \,
\sum_{p=1}^{M} \frac{2\,\Re{\rm e}[P_{p-1}(u)]}{p} \, \sum_{k=1}^{p-1} \,
R_{k}(M)
\, ,
\ea
with the factors of $\ft{1}{24} B_1^+$ and $-\ft{1}{4} \gamma_0^{+} \, \left[
(\gamma_0^{+})^2 + \alpha^{+} \right]$, respectively. Further, for the
completion of the last
three equations we need as well Eq.\ (\ref{M42})
for $L=2$ times $\ft{i}{8} \gamma_0^+$, and for $L = 0$ entering with
$\ft{i}{24}
B_1^+ - \ft{i}{4} \gamma_0^+ \, \left[ (\gamma_0^{+})^2 + \alpha^{+} \right]$.

Combining all these expressions and considering contributions proportional to
$Z_2(p)$, we
find that the infinite series present in separate terms cancel between the two
equations
(\ref{N4a1}) and Eq.\ (\ref{M42}) for $L=2$ and their net result is equal to
\be
- \ft{1}{16} \gamma_0^{+} \,
\sum_{p=1}^{M} \frac{2\,\Re{\rm e}[P_{p-1}(u)]}{p} \, Z_2(p)
\, .
\label{N4a8}
\ee
At the same time, the non-polynomial coefficients accompanying $Z_1(p)$ vanish
since the
leading order Baxter polynomial is an even function of the spectral parameter.
Turning
to the remaining two infinite-series contributions we deduce that they can be
resummed
into a concise expression such that the total inhomogeneity $U_3$ admits the
following
form with clearly separated polynomial terms
\ba
\label{N4b7ab}
U_3 [Q_0] =
&-& \ft{1}{32} \,  \sum_{p=1}^{M} \frac{2 \,\Re{\rm e}[P_{p-1}(u)]}{p} \,
\biggl\{
\, \left[ \, \ft{4}{3} B_1^{+} - 8 \gamma_0^{+}
\left( (\gamma_0^{+})^{2}
+
\alpha^{+} \right) \right] R_{p}(M)
\\
&+&
\sum_{k=1}^{p} R_{k}(M) \biggl[
\ft{1}{6} Z_3(p, k - 1)
-
2 \, \gamma_0^{+} \Big( \ft{1}{k} Z_1(p,k-1) - \ft12 Z_2(p,k)\Bigr)
\nonumber\\
&&\qquad\qquad\qquad\qquad\qquad
+
8 \gamma_0^{+} \left((\gamma_0^{+})^{2}+  \alpha^{+} \right) \biggr]
+
8 \Bigl( \gamma_0^{+}\alpha^{+} - 8 \beta^{+} \Bigr)
+ 2\gamma_0^{+} Z_2 (p)
\biggr\}
\nonumber\\
&+& \ft{i}{8} \left( \frac{1}{u^+} - \frac{1}{u^-} \right)
\left( \varepsilon^+ - 2 \gamma_0^+ \delta^+ \right)
\, . \!\!\!
\nonumber
\ea
Here we used the identity
\be
\frac{1}{L!}  \, Z_L(p,k-1)
\, = \,
\frac{1}{L!}  \, Z_L(p,k) + \frac{1}{(L-1)!}  \, \frac{Z_{L-1}(p,k-1)}{k}
\label{N4b8-}
\ee
to simplify intermediate results. Even though there are remaining non-polynomial
contributions,
(the last line in Eq.\ (\ref{N4b7ab})) that do not cancel on their own, they
will after we add
terms stemming from $U_2 [Q_1]$ as will be demonstrated in the next section
following the same
lines of reasoning as in Section \ref{npThreeloop} upon the replacement $Q_0(u)
\to Q_1(u)$.

%%%%%%%%%%%%%%%%%%%%%%%%%%%%%%%%%%%%%%%%%%%%%%%%%%%%%%%%%%%%%%%
\subsubsection{Inhomogeneity $U_2$}
%%%%%%%%%%%%%%%%%%%%%%%%%%%%%%%%%%%%%%%%%%%%%%%%%%%%%%%%%%%%%%%

To start with, we write the two-loop Baxter polynomial as
\be
Q_1 (u) = \sum_{p = 0}^M \Re{\rm e} [P_p (u)] R_p (M) \widetilde{r}_{1, p} (M)
\, ,
\label{NN1a+}
\ee
with (see Ref.\ \cite{KotRejZie08})
\be
\widetilde{r}_{1, p}(M) \, = \,
\widetilde{b} (M) + \gamma_0(M) \Big( S_1 (p + M)- S_1(M) - \ft{1}{2} \,  S_1(p)
\Big) - S_2(p) \, ,
\label{NN2+}
\ee
and
\be
\widetilde{b} (M) \, = \, b_1 (M) - S_1^2 (M) \,,
\label{NN3+}
\ee
where $b_1$ was introduced in Eq.\ (\ref{NormQ1Hahn}), such that $\widetilde{r}_{1,0}(M) =
Q_1 (\ft{i}{2})$.
Then we split $Q_1$ into a constant piece and the rest $\widetilde{Q}_1 (u)$
starting from
$P_{p > 0} (u)$
\be
\label{Q1split}
Q_1 (u) = \widetilde{r}_{1,0} (M)  + \widetilde{Q}_1 (u)
\, .
\ee
Then as in Section \ref{npThreeloop}, the inhomogeneity $U_2 [Q_1]$ is written
as a sum
of three terms: first, Eq.\ (\ref{N2}) multiplied by $\ft{i}4 \gamma_0^+$, where
we merely
replace $Q_0$ by $Q_1$ and $R_k$ by $R_{1,k}=R_k \widetilde{r}_{1, k}$; the difference of Eqs.\
(\ref{N3a})
and (\ref{N3a1}), both multiplied by $\ft{1}{16}$, with the same substitutions;
and Eq.\
(\ref{N3b}) accompanied by $\ft{i}{4} \gamma_0^+ \widetilde{r}_{1,0}$.

In order to perform the reduction of the inhomogeneity $U_2$ to a polynomial form, it is
sufficient to transform the summand of Eq.\ (\ref{NN1a+}) to a form involving just the
polynomial $P_p (u)$ itself rather than its real part, i.e., half the sum of $P_p (u)$ and
$P_p (- u)$. In Mellin space $P_p (u)$ corresponds to a polynomial in $z^n$ alone, but
not $(1-z)^n$. As can be observed easily, the terms proportional to $\widetilde{b} (M)$ and
$\gamma_0(M)$ are even functions of the spectral parameter and, as a consequence, do not
change after the substitution $u \to -u$. On the other hand, the term $\sim S_2(p)$ in the
right-hand side of (\ref{NN2+}) does not have this property. Indeed, from Appendix D we have
\be
\sum_{p = 0}^M P_p (- u) R_p (M)  S_2(p)
=
\sum_{p = 0}^M P_p (u) R_p (M) \Big[ V_{2,0}(p) - S_2(p) - 2 S_{- 2} (M) \Big]
\, ,
\label{NN3a+}
\ee
and deduce a complimentary representation of the two-loop Baxter polynomial
\be
Q_1 (u) = \sum_{p = 0}^M P_p (u) R_p (M) r_{1, p} (M)
\, ,
\label{NN1b+}
\ee
with
\be
r_{1, p}(M) \, = \,
\widetilde{b} (M) + \gamma_0(M) \Big( S_1 (p + M)- S_1(M) - \ft{1}{2} \,
S_1(p) \Big) - \ft{1}{2} \, V_{2,0}(p) + S_{- 2} (M)
\, .
\label{NN2b+}
\ee

Separating the infinite-series contributions from $U_2$ of the form as in Eq.\ (\ref{N4b7ab}),
we find that their coefficient conspire to give the same overall coefficient but with the
opposite sign, such that $U_2 [Q_1]$ can be cast in the form
\ba
\label{N4b7ac}
U_2 [Q_1]
&=&
\ft{1}{16} \,  \sum_{p=1}^{M} \frac{2 \,\Re{\rm e}[P_{p-1}(u)]}{p} \,
\biggl\{
\, \sum_{k=1}^{p} R_k (M) r_{1,k} (M) Z_1(p, k - 1)
\\
&&\qquad\qquad
- 4 \, \gamma_0^{+} R_p (M) r_{1,p} (M)
+ 2 \varepsilon^+
\biggr\}
- \ft{i}{8} \left( \frac{1}{u^+} - \frac{1}{u^-} \right)
\left( \varepsilon^+ - 2 \gamma_0^+ \delta^+ \right)
\, .
\nonumber
\ea
Therefore, the result for the sum of the two inhomogeneities $U_3 [Q_0] + U_2
[Q_1]$ is free from non-polynomialities and reads
\ba
\label{N4b7a}
&&- \ft{1}{16} \,  \sum_{p=1}^{M} \frac{\Re{\rm e}[P_{p-1}(u)]}{p} \,
\biggl\{
\, \left[ \, \ft{4}{3} B_1^{+} - 8 \gamma_0^{+} \left( (\gamma_0^{+})^{2}+
\alpha^{+} \right)
+
8 \gamma_0^{+} \, r_{1, p} (M) \right] \, R_{p}(M)
\\
\nonumber\\
&&\qquad\qquad\qquad
+
\sum_{k=1}^{p} R_{k}(M) \Bigl[
\ft{1}{6}  \, Z_3(p,k-1) - 2 \, \big( \ft{1}{k} \gamma_0^{+} + r_{1,k} (M) \big)
\, Z_1(p,k-1)
\nonumber\\
&&\qquad\qquad\qquad\qquad\qquad\qquad
+
\gamma_0^{+} \, Z_2(p,k)
+
8 \gamma_0^{+} \left((\gamma_0^{+})^{2}+  \alpha^{+} \right) \Bigr]
- 16 C_{3,0} + 2\gamma_0^{+} Z_2 (p)
\biggr\}
\, , \!\!\!
\nonumber
\ea
with $C_{3,0}$ combining the functions $C_{3,0}
 =  4  \beta^{+} + \sfrac{1}{4}  \varepsilon^{+} - \sfrac{1}{2}
\gamma_0^{+} \alpha^{+}$. %\label{N4b6}
The corresponding contribution to $Q_3$ is then given by
\ba
Q_{3}^{(npp)}(u)
\, = \,
\sum_{k=0}^{M} 2 \, \Re{\rm e}[ P_{k}(u)] \, R_{k} (M) r_{3,k}^{(npp)}(M) \, \,
,
\label{N4b8}
\ea
where we give $r_{3,k}^{(npp)}$ in the same order as the corresponding inhomogeneities
appear in Eq.~\eqref{N4b7a}
\ba
\label{N4b9}
r_{3,k}^{(npp)}(M)
&=&
-\Big[ \ft{1}{24}  \, B_1^{+} -\ft{1}{4} \,
\gamma_0^{+} \left( (\gamma_0^{+})^{2}+  \alpha^{+} \right)\Big] S_{3}(k)
-\ft{1}{4} \, \gamma_0^{+} \, \widehat{S}_{3,1}(k)-
\ft{1}{192}  \,  V_{3,3}(k)\nonumber\\
&& +\ft{1}{16} \, \gamma_0^{+} \, \widehat{V}_{3,2}(k) + \ft{1}{16} \,  V_{3,1,1}(k)
-\ft{1}{32} \, \gamma_0^{+} \, \widetilde{V}_{3,2}(k) \nonumber\\
&&-\ft{1}{4}\,
\gamma_0^{+} \left( (\gamma_0^{+})^{2}+  \alpha^{+} \right)\widetilde{V}_{3,0}(k)
+ \ft12 \, C_{3,0} \, W_{3}(k)-\ft{1}{16} \gamma_0^{+} \, W_{3,2}(k)
\, .
\ea
It is expressed via the following nested sums
\ba\label{N4b10}
\widehat{S}_{a,b}(k)
&=&
\sum_{l=1}^{k} \, \frac{1}{l^a} \, r_{b,l}(M)
\, , \\
W_{a,b}(k)
&=&
\sum_{l=1}^{k} \, \frac{1}{l^a \, R_l(M)} \, Z_b(l)
\, , \\
\widehat{V}_{a,b,c}(k)
&=&
\sum_{r=1}^{k} \, \frac{1}{r^a \, R_r(M)}
\, \sum_{m=1}^{r} \, \frac{R_{c,m}(M)}{m} \, Z_{b-1}(r,m-1)
\, , \\
\widetilde{V}_{a,b,c}(k)
&=&
\sum_{r=1}^{k} \, \frac{1}{r^a \, R_r(M)} \, \sum_{m=1}^{r} \, R_{c, m}(M) \,
Z_b (r,m)
\, , \\
V_{a,b,c}(k)
&=&
\sum_{r=1}^{k} \, \frac{1}{r^a \, R_r(M)} \, \sum_{m=1}^{r} \, R_{c,m}(M) \,
Z_b (r,m-1)
\, .
\ea
These are related to the previously introduced sums (\ref{N8}) -- (\ref{N8last}) via
\be
W_{a}(k) \, = \, W_{a,0}(k)
\, , \quad
\widehat{V}_{a,b} (k) \, = \, \widehat{V}_{a,b,0} (k)
\, , \quad
\widetilde{V}_{a,b} (k) \, = \,  \widetilde{V}_{a,b,0} (k)
\, , \quad
V_{a,b} (k) \, = \, V_{a,b,0} (k)
\, .
\ee

%%%%%%%%%%%%%%%%%%%%%%%%%%%%%%%%%%%%%%%%%%%%%%%%%%%%%%%%%%%%%%%
\subsubsection{Inhomogeneity $U_1$}
%%%%%%%%%%%%%%%%%%%%%%%%%%%%%%%%%%%%%%%%%%%%%%%%%%%%%%%%%%%%%%%

Finally, we turn to the polynomial inhomogeneity (\ref{N4c1}) of the non-polynomial part
of the three-loop Baxter function $Q_2^{(np)}$. It can be represented as a sum
\be
\label{N4c2}
U_1 [Q_2^{(np)}] = U_1^{[0]} [Q_2^{(np)}] + U_1^{[1]} [Q_2^{(np)}] \, ,
\ee
with
\ba
U_1^{[0]} [Q_2^{(np)}]
&=&
\ft{1}{2} \, \left(  Q_2^{(np)} (u + i)+  Q_2^{(np)} (u-i) - 2
Q_2^{(np)} (u)\right)
\, ,
\label{N4c3}  \\
U_1^{[1]} [Q_2^{(np)}] &=&
- \gamma_0^+ \left( i u^+ Q_2^{(np)} (u+i) - i u^- Q_2^{(np)} (u-i)
+ (2M + 1) Q_2^{(np)} (u)\right)
\, .
\label{N4c4}
\ea
{}From this, it is immediate to find the corresponding contribution to $Q_3$ to be
\be
\label{N4c5}
Q_3^{(pnp),[i]}(u) =
\sum_{k=0}^{M}
\Re{\rm e}[ P_{k}(u)] \, R_k(M) \, r_{3,k}^{(pnp),[i]} (M)
,
\qquad
(i=0,1)
\, ,
\ee
where
\ba
r_{3,k}^{(pnp),[0]}(M)
&=& \widehat{R}_{2,M}^{(np)} (M) \left(\widetilde{b}(M)+W_2(k)\right)
- \sum_{m=1}^k \, \frac{\widehat{R}_{2,m}^{(np)} (M)}{m^2 R_m(M)}
\, , \label{N4d7} \\
r_{3,k}^{(pnp),[1]}(M)
&=& 2 \gamma_0^+
\,\sum_{p=1}^{k} \,
\left\{\frac{2}{p + M} \, r_{2,p-1}^{(np)}(M)-
\frac{1}{p}  r_{2,p}^{(np)} (M)
\right\}
\, , \label{N4c7}
\ea
and
\be
\widehat{R}_{2,m}^{(np)} (M) \, = \, \sum_{p=1}^{m} \, R_{2,p}^{(np)} (M)
\, , \quad
R_{2,p}^{(np)}(M)=R_p(M)r_{2,p}^{(np)}(M)
\, , \quad
\widehat{R}_{2,M}^{(np)} (M) \, = \, Q_2^{(np)} (-\ft{i}{2})
\, .  \label{N4e7}
\ee

%%%%%%%%%%%%%%%%%%%%%%%%%%%%%%%%%%%%%%%%%%%%%%%%%%%%%%%%%%%%%%%
\section{Anomalous dimensions}
%%%%%%%%%%%%%%%%%%%%%%%%%%%%%%%%%%%%%%%%%%%%%%%%%%%%%%%%%%%%%%%

Making use of the explicit solution to the Baxter equation to four loops, we can
immediately calculate its derivatives at the argument $u = \pm \ft{i}{2}$ (see, e.g.,
Appendix \ref{App:Derivatives}) and find the corresponding anomalous dimensions by
means of Eq.\ (\ref{AllOrderAD}) expanded to the required order of perturbation theory.
The results are\footnote{All results are given for even values of $M$. An analytical 
continuation to complex values can be found in \cite{Kotikov:2005gr}.}
\ba
\gamma_0
&=&
2 S_1
\, , \\
\gamma_1
&=&
- S_3 - S_{- 3}
+ 2 S_{-2,1} - 2 S_1 \left( S_2 + S_{- 2} \right)
\, , \\
\gamma_2
&=&
S_5 + 3 S_{- 5} - 2 S_{- 3} S_2 + 2 S_{- 2} S_3 - 24 S_{- 2,1,1,1}
- 6 S_{- 4, 1} - 6 S_{- 3, 2} - 6 S_{-2, 3}
\nonumber\\
&+& 12 S_{- 3,1,1} + 12 S_{- 2,1,2} + 12 S_{- 2,2,1}
+ \left( S_2 + 2 S_1^2 \right) \left( 3 S_{- 3} + S_3 - 2 S_{- 2,1} \right)
\nonumber\\
&+&
S_1
\left(
8 S_{- 4} + S_{- 2}^2 + 4 S_2 S_{- 2} + 2 S_2^2 + 3 S_4
- 12 S_{- 3,1} - 10 S_{- 2,2} + 16 S_{- 2,1,1}
\right)
\, , \\
\gamma_3^{(\rm asy)}
&=&
{4\, S_{-7}+6\, S_{7}}+ 2\,( S_{-3,1,3} + S_{-3,2,2} +
      S_{-3,3,1} + S_{-2,4,1} )
  + 3\,( -S_{-2,5}\nonumber\\
  &+& S_{-2,3,-2} ) +4\,( S_{-2,1,4}
- S_{-2,-2,-2,1} -
      S_{-2,1,2,-2} - S_{-2,2,1,-2} -
      S_{1,-2,1,3} \nonumber \\
&-& S_{1,-2,2,2} -
      S_{1,-2,3,1} )
  + 5\,( -S_{-3,4}
+ S_{-2,-2,-3} )
  + 6\,(- S_{5,-2} \nonumber\\
&+& S_{1,-2,4} -
      S_{-2,-2,1,-2} - S_{1,-2,-2,-2} )
  + 7\,( -S_{-2,-5}
+ S_{-3,-2,-2} \nonumber\\
&+&
      S_{-2,-3,-2} + S_{-2,-2,3} )
  + 8\,( S_{-4,1,2} + S_{-4,2,1} - S_{-5,-2} - S_{-4,3}\nonumber \\  &-&
      S_{-2,1,-2,-2}
+ S_{1,-2,1,1,-2} )
  + 9\,S_{3,-2,-2}
  -10\,S_{1,-2,2,-2}
  + 11\,S_{-3,2,-2}\nonumber\\
 &+& 12\,( -S_{-6,1} + S_{-2,2,-3}
+ S_{1,4,-2} + \!S_{4,-2,1} + \!
      S_{4,1,-2} - \!S_{-3,1,1,-2} -\!
      S_{-2,2,-2,1}\nonumber\\
&-& \!S_{1,1,2,3} - \!
      S_{1,1,3,-2} - \!S_{1,1,3,2} - \!
      S_{1,2,1,3}
- S_{1,2,2,-2} -
      S_{1,2,2,2} - S_{1,2,3,1} -
      S_{1,3,1,-2} \nonumber \\
&-& S_{1,3,1,2} -
      S_{1,3,2,1} - S_{2,-2,1,2} -
      S_{2,-2,2,1} - S_{2,1,1,3}
-
      S_{2,1,2,-2} - S_{2,1,2,2}\nonumber\\
&-&
      S_{2,1,3,1} - S_{2,2,1,-2} -
      S_{2,2,1,2} - S_{2,2,2,1} -
      S_{2,3,1,1} - S_{3,1,1,-2} -
      S_{3,1,1,2}
- S_{3,1,2,1}\nonumber\\
&-& S_{3,2,1,1} )
+ 13\,S_{2,-2,3}
  -14\,S_{2,-2,1,-2}
  + 15\,( S_{2,3,-2} + S_{3,2,-2} )
 \nonumber \\
&+& 16\,( S_{-4,1,-2}
+ S_{-2,1,-4} - \!
      S_{-2,-2,1,2} - \!S_{-2,-2,2,1} - \!
      S_{-2,1,-2,2} - \!S_{-2,1,1,-3} \nonumber\\
&-& \!
      S_{1,-3,1,2} - \!S_{1,-3,2,1} - \!
      S_{1,-2,-2,2}
- S_{2,-2,-2,1} +
      S_{-2,1,1,-2,1} + S_{1,1,-2,1,-2}\nonumber \\
 &+&
      S_{1,1,-2,1,2} + S_{1,1,-2,2,1} )
  -17\,S_{-5,2}
  + 18\,( -S_{4,-3}
- S_{6,1} + S_{1,-3,3} )\nonumber\\
  &+& 20\,( -S_{1,-6} - S_{1,6} -
      S_{4,3} + S_{-5,1,1} +
      S_{-4,-2,1} + S_{-3,-2,2} +
      S_{-2,-4,1} \nonumber\\
&+& S_{-2,-3,2} +
      S_{1,3,3} + S_{3,1,3} +
      S_{3,3,1} - S_{1,1,-2,3} -
      S_{1,2,-2,-2} - S_{2,1,-2,-2} )\nonumber \\
 &-&21\,S_{3,4}
+ 22\,( S_{1,-2,-4} + S_{2,2,3} +
      S_{2,3,2} + S_{3,-2,2} + S_{3,2,2})
+ 23\,( -S_{-3,-4} \nonumber \\
&-& S_{5,2} +
      S_{2,-2,-3} )
+ 24\,( -S_{-4,-3} + S_{1,-4,-2} -
      S_{1,-3,1,-2} - S_{1,1,1,4} -
      S_{1,1,4,1}\nonumber\\
&-& S_{1,3,-2,1} -
      S_{1,4,1,1} - S_{3,-2,1,1}
- S_{3,1,-2,1} - S_{4,1,1,1} +
      S_{-2,-2,1,1,1} + S_{-2,1,-2,1,1}\nonumber\\
&+&
      S_{1,-2,-2,1,1} + S_{1,-2,1,-2,1} +
      S_{1,1,-2,-2,1}
+ S_{1,1,1,-2,-2} +
      S_{1,1,2,-2,1} + S_{1,2,1,-2,1}\nonumber\\
&+&
      S_{2,1,1,-2,1} )
+ 25\,S_{2,-3,-2}
  + 26\,( -S_{2,5} + S_{1,4,2}
+ S_{2,4,1} + S_{4,1,2} + S_{4,2,1})\nonumber\\
&+& 28\,( S_{1,2,4} + S_{2,1,4} -
      S_{-3,1,-2,1} - S_{-2,1,-3,1} -
      S_{1,-2,1,-3} )
+ 30\,S_{-3,1,-3} \nonumber\\
  &+& 32\,( S_{1,5,1} + S_{5,1,1} -
      S_{-3,-2,1,1} - S_{-2,-3,1,1} -
      S_{1,-3,-2,1} - S_{1,-2,-3,1} \nonumber\\
&-&S_{2,2,-2,1} + S_{1,2,-2,1,1} +
      S_{2,1,-2,1,1} - S_{1,1,1,-2,1,1} )
  + 36\,( S_{1,1,5} + S_{1,3,-3} \nonumber\\
&+&      S_{3,1,-3}
- S_{1,1,-3,-2} - \!
      S_{1,1,-2,-3} - \!S_{1,1,2,-3} - \!
      S_{1,2,-2,2} - \!S_{1,2,1,-3} - \!
      S_{2,1,-2,2} \nonumber \\
&-& \!S_{2,1,1,-3} )
  + \!38\,S_{-3,-3,1}
+ 40\,( -S_{1,-4,1,1} - S_{2,-3,1,1} +
      S_{1,1,1,-2,2} )\nonumber\\
&-&41\,S_{3,-4}
  + 42\,( -S_{2,-5} + S_{1,-4,2} +
      S_{1,-3,-3} )
+ 44\,( S_{1,-5,1} + S_{2,-3,2}
+      S_{3,-3,1} )\nonumber \\
&+& 46\,S_{2,2,-3}
  + 48\,S_{1,1,-3,1,1}
  + 60\,( S_{1,1,-5} - S_{1,1,-3,2} )
+ 62\,S_{2,-4,1}
+ 64\,S_{1,1,1,-3,1}\nonumber\\
 &+& 68\,( S_{1,2,-4} + S_{2,1,-4} -
      S_{1,2,-3,1}- S_{2,1,-3,1} )
-72\,S_{1,1,1,-4}
-80\,S_{1,1,-4,1}\nonumber\\
&-&\zeta_3 S_1(S_3-S_{-3}+2\,S_{-2,1})
\, .
\ea
These agree with expressions found from explicit calculations of Feynman
diagrams at one
\cite{Lip98} and two loops \cite{KotLipVel03}, and three \cite{KotLipOniVel04}
and four-loop
\cite{KotLipRejStaVel07} results obtained with the use of numerical solution of
Bethe equations
and the principle of maximal transcendentality.

%%%%%%%%%%%%%%%%%%%%%%%%%%%%%%%%%%%%%%%%%%%%%%%%%%%%%%%%%%%%%%%
\section{Five-loop dressing and reciprocity}
\label{five-loop dresci}
%%%%%%%%%%%%%%%%%%%%%%%%%%%%%%%%%%%%%%%%%%%%%%%%%%%%%%%%%%%%%%%

Finally, let us partially address the five loop order, namely, the one stemming
from the
dressing phase $\Theta (u)$. The dressing part of the five-loop Baxter
polynomial can be
written in terms of contributions with decreasing transcendentality as
\be
Q^{(d)}_4 (u)
=
\zeta_3 \, \left( Q_{\zeta_3}^{(p)} (u) + Q_{\zeta_3}^{(np)} \right) + \zeta_5
\, Q_{\zeta_5}^{(p)} (u)
\, ,
\ee
where we decomposed the term accompanying $\zeta_3$ according to the
nomenclature
of polynomial and non-polynomial inhomogeneities. Their calculation in the
Wilson basis echoes
the one performed in the previous section and yields for polynomial
contributions,
\ba
Q_{\zeta_5}^{(p)} (u)
&=&
a_{4, \zeta_5} \, Q_0 + \ft{5}{16} \, S_1 \, T_{(0,1)}
\, , \\
Q_{\zeta_3}^{(p)} (u)
&=&
a_{4, \zeta_3}^{(p)}  \, Q_0
+ c_{(1,0)}  \, T_{(1,0)}
+ c_{(2,0)}  \, T_{(2,0)}
+ c_{(0,1)}  \, T_{(0,1)}
+
c_{(1,1)} \, T_{(1,1)}
+
c_{(0,2)} \, T_{(0,2)}
\, . \quad
\ea
Here
\ba
c_{(1,0)}
&=&
- \ft{1}{2} S_1^4 - \ft{1}{2} S_2 S_1^2 + S_{1,1} S_1^2
\, , \\
c_{(2,0)}
&=&
\ft{1}{4} S_1^3 - \ft{1}{4} S_{-2} S_1
\, , \nonumber\\
c_{(0,1)}
&=&
- \ft{5}{8} S_1^3 + \ft{1}{4} \widetilde{S}_1 S_1^2 - \ft{1}{4} S_2 S_1
- \ft{1}{16} S_{-3} - \ft{1}{16} S_3
+
\ft{1}{8} S_{1,-2} + \ft{1}{8} S_{1,2} + \ft{1}{8} S_{2,1}
\, , \nonumber\\
c_{(1,1)} &=& - \ft{1}{8} S_1^2
\, , \nonumber\\
c_{(0,2)} &=& \ft{1}{96} S_1
\, , \nonumber
\ea
and the degree-reducing constants are
\ba
a_{4, \zeta_5}
&=&
-\tfrac{5}{4} S_1 \left(S_{-2}+S_2\right)
\, , \\
a_{4, \zeta_3}
&=&
-\tfrac{15}{4}S_1^5+4 {\widetilde{S}}_1 S_1^4-{\widetilde{S}}_1^2
S_1^3+\tfrac{13}{4} S_{-2} S_1^3-\tfrac{7}{4} S_2 S_1^3+
{\widetilde{S}}_2 S_1^3+3 S_{1,1} S_1^3-S_{-3} S_1^2 +\nonumber\\
&&- S_3 S_1^2-3 S_{-2}{\widetilde{S}}_1 S_1^2+
S_2 {\widetilde{S}}_1 S_1^2-2 {\widetilde{S}}_1 S_{1,1} S_1^2-\tfrac{1}{2}
S_{-2}^2  S_1+\tfrac{1}{2} S_2^2 S_1 +S_{-2} {\widetilde{S}}_1^2 S_1 +\nonumber\\
&&- S_{-4} S_1+\tfrac{5}{4} S_{-2} S_2 S_1-S_4 S_1-S_{-2} {\widetilde{S}}_2
S_1+\tfrac{1}{4} S_{-3} S_{-2}+\tfrac{1}{4} S_{-3} S_2+\tfrac{1}{4} S_{-2}
S_3+\nonumber\\
&&+ \tfrac{1}{4}S_2 S_3-\tfrac{1}{2} S_{-2}
   S_{1,-2} - \tfrac{1}{2} S_2 S_{1,-2}-\tfrac{1}{2} S_{-2} S_{1,2}-\tfrac{1}{2} S_2 S_{1,2}-
\tfrac{1}{2} S_{-2} S_{2,1}-\tfrac{1}{2} S_2 S_{2,1}\, .\nonumber
\ea
Finally, the non-polynomial term, obeys the equation
\be
\mathfrak{B}[Q_{\zeta_3}^{(np)}] = -i\,S_1^2\, U_{\zeta_3}[Q_0],
\ee
with
\ba
U_{\zeta_3}[Q_0] &=& \frac{Q_0(u+i)-Q_0(u)}{u^+}-\frac{Q_0(u-i)-Q_0(u)}{u^-}
\\
&=& 2\,i\,\sum_{p=1}^M\frac{\Re{\rm e}[ P_{p-1}](u)}{p}\sum_{k=1}^p R_k(M)~.
\nonumber
\ea
The polynomial $Q_{\zeta_3}^{(np)}$ can be computed according to the method
spelled out above. It reads
\be
Q_{\zeta_3}^{(np)} (u)
=
2 S_1^2 \, \sum_{p = 0}^M \Re{\rm e} [P_p (u)] \sum_{l = 1}^p \frac{1}{l^3} \frac{R_p (M)}{R_l (M)}
\sum_{k = 1}^l R_k (M)
\, .
\ee

Substituting these findings in Eq.\ (\ref{AllOrderAD}) expanded to fifth order
in the 't Hooft coupling,
we find the dressing part of the five-loop anomalous dimensions of twist-two
operators,
\be
\gamma_5^{(\rm asy)} = \dots + \zeta_3 \, \gamma_5^{\zeta_3} + \zeta_5 \, \gamma_5^{\zeta_5}
\, ,
\ee
with spin-dependent functions $\gamma_5^{\zeta_3}$ and $\gamma_5^{\zeta_5}$
obeying the principle of maximal
transcendentality (as well as the absence of $-1$ indices in nested harmonic
sums)
\ba
\gamma_5^{\zeta_5}
&=&
\tfrac{5}{2} S_{-4} - \tfrac{5}{2} S_4 - \tfrac{15}{2} S_{-3,1} - 5 S_{-2,2} -
\tfrac{5}{2} S_{1,-3}
+ \tfrac{5}{2} S_{1,3} + \tfrac{5}{2} S_{3,1} + \nonumber\\
&& +  10 S_{-2,1,1} + 5 S_{1,-2,1}
\, , \\
\gamma_5^{\zeta_3} &=&
-\,S_{-6}+\,S_6+11 \,S_{-5,1}+4 \,S_{-4,-2}+20 \,S_{-4,2}+9 \,S_{-3,-3}+14
\,S_{-3,3}+\nonumber\\
&& + 4 \,S_{-2,-4}+4 \,S_{-2,4}
 + 3 \,S_{1,-5}-3 \,S_{1,5}+4 \,S_{2,-4}-4 \,S_{2,4}-3 \,S_{3,3}-2 \,S_{4,-2} +
\nonumber\\
&& -4
   \,S_{4,2}-3 \,S_{5,1}-32 \,S_{-4,1,1}
 -8 \,S_{-3,-2,1}-10 \,S_{-3,1,-2}-28 \,S_{-3,1,2}-28 \,S_{-3,2,1} + \nonumber\\
&& -6 \,S_{-2,-3,1}-2 \,S_{-2,-2,2}-10 \,S_{-2,1,-3}
 -10 \,S_{-2,1,3}-6
   \,S_{-2,2,-2}-12 \,S_{-2,2,2}+\nonumber\\
&& -10 \,S_{-2,3,1}-20 \,S_{1,-4,1}-6 \,S_{1,-3,-2}-24 \,S_{1,-3,2}
 -8 \,S_{1,-2,-3}-10 \,S_{1,-2,3}-4 \,S_{1,1,-4}+\nonumber\\
&& + 4 \,S_{1,1,4}-2 \,S_{1,2,-3}+2 \,S_{1,2,3}+2
   \,S_{1,3,-2}+2 \,S_{1,3,2}
 + 2 \,S_{1,4,1}-14 \,S_{2,-3,1}-2 \,S_{2,-2,-2}+\nonumber\\
&& -10 \,S_{2,-2,2}-2 \,S_{2,1,-3}+2 \,S_{2,1,3}+2 \,S_{2,3,1}-2 \,S_{3,-2,1}
 + 2 \,S_{3,1,-2}+2 \,S_{3,1,2}+2
   \,S_{3,2,1}+\nonumber\\
&&  + 4 \,S_{4,1,1}+36 \,S_{-3,1,1,1}+4 \,S_{-2,-2,1,1}+8 \,S_{-2,1,-2,1}+
 + 12 \,S_{-2,1,1,-2}+12 \,S_{-2,1,1,2}+\nonumber\\
&& + 12 \,S_{-2,1,2,1}+12 \,S_{-2,2,1,1}+36 \,S_{1,-3,1,1}+8
   \,S_{1,-2,-2,1}
 + 8 \,S_{1,-2,1,-2}+20 \,S_{1,-2,1,2}+\nonumber\\
&& + 20 \,S_{1,-2,2,1}+16 \,S_{1,1,-3,1}+4 \,S_{1,1,-2,-2}+12 \,S_{1,1,-2,2}
+ 4 \,S_{1,2,-2,1}+16 \,S_{2,-2,1,1}+\nonumber\\
&& + 4 \,S_{2,1,-2,1} -24  \,S_{1,-2,1,1,1}-16 \,S_{1,1,-2,1,1} \, .
\ea

%%%%%%%%%%%%%%%%%%%%%%%%%%%%%%%%%%%%%%%%%%%%%%%%%%%%%%%%%%%%%%%
\subsection{Parity invariance}
%%%%%%%%%%%%%%%%%%%%%%%%%%%%%%%%%%%%%%%%%%%%%%%%%%%%%%%%%%%%%%%

As we have seen in Section 2, the conserved charge $\mathfrak{Q}_2$ acquires
perturbative corrections
which shift the bare total conformal spin of Wilson
operators by their anomalous dimension $\gamma (g)$ to the renormalized
one,
\be
j_0 = M + 1 \qquad \to \qquad j = M + 1 + \ft12 \gamma (g)
\, .
\ee
This phenomenon implies that the anomalous dimensions can be defined more
naturally as functions of
the renormalized rather than bare spin, such that one can define a new function
$P$ of argument $J$,
\be
\label{RRelation}
\gamma = P (M + \ft12 \gamma)
\, .
\ee
The parity invariance property of anomalous dimensions is then formulated as
invariance
of $P (j)$ under the reflection map $J \to - J$ with $J^2 = j_0 (j_0 + 1)$
\cite{DokMar06,BasKor06}. This condition results in an infinite number of
relations for coefficients accompanying odd powers of the Lorentz spin in the
large-$M$
expansion in terms of corresponding even powers. For finite $M$ this property
gets translated
into the presence of parity-even combinations of nested harmonic sums $\Omega_{k_1,
k_2, \dots}$
only, which can have positive even and negative odd $k_i$'s
\cite{BecFor09,ForBec09}.
These functions are defined as follows. For a single-index harmonic sum, they do
coincide
with usual harmonic numbers, while for more than one index, they are defined
recursively,
\be
\Omega_{k_1} = S_{k_1}
\, , \qquad
\Omega_{k_1, k_2} = \omega_{k_1} (\Omega_{k_2})
\, , \qquad
\Omega_{k_1, k_2, k_3} = \omega_{k_1} (\Omega_{k_2, k_3})
\, , \dots ,
\ee
with the involved map defined by
\be
\omega_{k_1} ( S_{k_2, k_3, \dots})
=
S_{k_1, k_2, k_3, \dots} - \ft12 S_{{\rm sign}(k_1){\rm sign}(k_2) (|k_1| +
|k_2|), k_3 \dots}
\, .
\ee

Let us establish the parity invariance property for the dressing part of the
five-loop twist-two
anomalous dimensions we determined above. Expanding Eq.\ (\ref{RRelation}) in the 't
Hooft coupling
and taking into account that dressing appears firstly at four loops, we find the
five loop dressing
contribution to $P(M)$
\be
\widehat{P}_5
=
\widehat{\gamma}_5
-
\tfrac{1}{4}\widehat{\gamma}_4\,\gamma_1'
-
\tfrac{1}{2}\widehat{\gamma}_4'\,\gamma_1,
\ee
where the hat on symbols denotes that we consider only their dressing parts.
Explicitly, we use
\ba
\gamma_1 &=& 2\,S_1, \\
\widehat{\gamma}_4 &=& S_1\,(S_{-3}-S_3-2\,S_{-2,1})\,\zeta_3, \\
\widehat{\gamma}_5 &=& \zeta_3 \, \gamma_5^{\zeta_3} + \zeta_5 \,
\gamma_5^{\zeta_5}
\, .
\ea
A long but straightforward calculation gives
\be
{\widehat P}_5 = \zeta_5 {\widehat P}_5^{\zeta_5} + \zeta_3 {\widehat
P}_5^{\zeta_3}
\, ,
\ee
where
\ba
{\widehat P}_5^{\zeta_5} &=& -\tfrac{5}{4} \left(\Omega _{-4}-2 \Omega _{1,3}-2
\Omega _{3,1}-8 \Omega _{-2,1,1}-4 \Omega _{1,-2,1}\right), \nonumber\\
{\widehat P}_5^{\zeta_3} &=& -\tfrac{5}{2} \Omega _{-6} -2 \Omega _{-4,-2}-2
\Omega _{-2,-4}+\Omega _{3,3}+\nonumber\\
&& + 2 \left(4 \Omega _{-4,1,1}+\Omega _{-2,1,3}+\Omega _{-2,3,1}+4 \Omega
   _{1,-4,1}+2 \Omega _{1,-2,3}+3 \Omega _{1,1,-4}+\Omega _{1,3,-2}+\right.
\nonumber\\
&& \left. +\Omega _{3,-2,1}+\Omega _{3,1,-2}\right)+4 \left(\Omega _{-2,-
2,1,1}+2 \Omega
   _{-2,1,-2,1}+3 \Omega _{-2,1,1,-2}+2 \Omega _{1,-2,-2,1}+\right. \nonumber \\
&& \left. + 2 \Omega _{1,-2,1,-2}+\Omega _{1,1,-2,-2}\right)-8 \left(3 \Omega
_{1,-2,1,1,1}+2 \Omega
   _{1,1,-2,1,1}\right) +\nonumber\\
&& +  2 \zeta_2
\left(
-  \Omega _{-4}
+ \left(\Omega _{1,3} + \Omega _{3,1} \right)
+
2 \left(3 \Omega _{-2,1,1} + 2 \Omega _{1,-2,1} + \Omega_{1,1,-2}\right)
\right)+
\nonumber\\
&& + 7\,\zeta_4\,\Omega _{1,1}
\, ,
\ea
which indeed obeys the parity-invariance properties spelled out at the beginning
of the section.

%%%%%%%%%%%%%%%%%%%%%%%%%%%%%%%%%%%%%%%%%%%%%%%%%%%%%%%%%%%%%%%
\section{Conclusions}
%%%%%%%%%%%%%%%%%%%%%%%%%%%%%%%%%%%%%%%%%%%%%%%%%%%%%%%%%%%%%%%

In this work we have developed an improved formalism for the analytical solution
of the multiloop
Baxter equation. As a demonstration of the efficiency of the framework, we found
the four-loop
Baxter polynomial and derived in a completely analytical form the resulting
anomalous dimensions.
This was possible largely due to an improved treatment of superficially
non-polynomial terms in
the Baxter equation. While in the previous consideration, the latter yielded
multiple sums with
Stirling numbers of the first and second kind involved and, as a consequence,
hampering
straightforward analytical calculation of derivatives of the Baxter function at
fixed points
which enter the definition of anomalous dimensions. They were treated making use
of the principle
of maximal transcendentality \cite{Kotikov:2002ab}
by writing down the most general expression in
terms of nested
harmonic sums and then fitting the multiplicative rational coefficients to
numerical data. In
the current analysis this difficulty was overcome.

Next we found a more concise representation of the polynomial contribution to
Baxter function
by using the basis of Wilson rather than continuous Hahn polynomials. The
complexity of these
expressions was reduced roughly in half. Nevertheless it should be stated that
our analysis of the non-polynomial terms is still in favor of a representation
in continuous Hahn polynomials. It might still be interesting to completely
restrict also the non-polynomial terms to a Wilson basis. As such it would be possible
to compare one-to-one the analytic properties of the Baxter function of twist-two
and -three operators, in order to pin down the origin of the asymptotic character
of the Baxter equation as well as the Bethe ansatz.

Finally, we provided further evidence towards parity invariance of twist-two
multiloop anomalous
dimensions by calculating the dressing contribution to the five-loop result and
showing that they
obey the same theorem as was established earlier.

What we did not address are the wrapping effects in twist-two operators, which
emerge starting from
four loops. The latter are known to be described by a generalized L\"uscher
formula which reads for
the case of twist-$L$ operators \cite{BajJanLuk08},
\be
\label{WrappingGamma}
\gamma^{(\rm wrap)} (g)
=
- i g^4 \gamma_0^2
\sum_{n = 1}^\infty \res\limits_{z = i n}
\left(\frac{g^2}{z^2 + n^2} \right)^L \frac{T^2 (z, n)}{R (z, n)}
+
\mathcal{O} \left( g^{2 (L + 3)} \right)
\, .
\ee
It is written in terms of
\be
R (z, n) =
Q_0 \left( \ft12 z - \ft{i}{2} (n - 1) \right)
Q_0 \left( \ft12 z + \ft{i}{2} (n - 1) \right)
Q_0 \left( \ft12 z + \ft{i}{2} (n + 1) \right)
Q_0 \left( \ft12 z - \ft{i}{2} (n + 1) \right)
\, ,
\ee
and the function
\be
T (z, n) = \sum_{m = 0}^{n - 1}
\frac{
Q_0 \left( \ft12 z - \ft{i}{2} (n - 1) + i m \right)
}{
\left[ \left( m - \ft12 n \right) - \ft{i}{2} z \right]
\left[ \left( m + 1 - \ft12 n \right) - \ft{i}{2} z \right]
}
\, .
\ee
These correspond to one-loop corrections in a sigma model. The explicit formula
for
twist-two operators was found in Ref.~\cite{BajJanLuk08} and reads
\be
\gamma^{(\rm wrap)}_3
=
S_1^2
\left(
- 5 \zeta_5 - 4 S_{- 2} \zeta_3
- 2 S_5 + 2 S_{- 5} + 4 S_{4,1} - 4 S_{3,- 2} + 4 S_{- 2, - 3} - 8 S_{- 2, -
2,1}
\right)
\, .
\ee

Our method can be viewed as a first step towards the analytical computation of
the five-loop twist-two
anomalous dimensions. Wrapping contributions are not included here. However the
knowledge of the
asymptotic prediction derived with it will allow one to analyze its analytical
structure in the
complex spin $M$ plane and constrain potential wrapping structures. These
questions will be
addressed elsewhere.

\vspace{5mm}

\noindent This work was supported by the U.S. National Science Foundation under
grant no.\
PHY-0757394 (A.B.) and by the Russian Foundation for Basic Research through
Grant No. 07-02-00902-a (A.K.). 
S.Z.~thanks the DAAD for a short term lectureship and the
University of
Herat in Afghanistan for kind hospitality. S.Z.~also thanks Matthias Staudacher
for discussions.
A.K.~thanks the Albert-Einstein-Institut for kind hospitality and for a
financial support.
%%%%%%%%%%%%%%%%%%%%%%%%%%%%%%%%%%%%%%%%%%%%%%%%%%%%%%%%%%%%%%%
\appendix
%%%%%%%%%%%%%%%%%%%%%%%%%%%%%%%%%%%%%%%%%%%%%%%%%%%%%%%%%%%%%%%
\section{Inhomogeneities}
\label{App:Inhomogeneities}
%%%%%%%%%%%%%%%%%%%%%%%%%%%%%%%%%%%%%%%%%%%%%%%%%%%%%%%%%%%%%%%

The inhomogeneities of the four-loop expansion of the Baxter equation appear
with the following
multiplicative functions
\ba
\alpha^+
&=&
\ft14
\frac{Q_0''(\tfrac{i}{2})}{Q_0 (\tfrac{i}{2})}
\, , \\
\beta^+
&=&
\ft{i}{192}
\frac{Q_0'''(\tfrac{i}{2})}{Q_0 (\tfrac{i}{2})}
\, , \nonumber\\
\chi^+
&=&
- \ft{1}{16} \frac{Q_1 (\tfrac{i}{2}) Q_0''(\tfrac{i}{2})}{Q_0^2 (\tfrac{i}{2})}
+ \ft{1}{16} \frac{Q_1'' (\tfrac{i}{2})}{Q_0 (\tfrac{i}{2})}
+ \ft{1}{192} \frac{Q_0^{(4)} (\tfrac{i}{2})}{Q_0 (\tfrac{i}{2})}
\, , \nonumber\\
\delta^+
&=&
\frac{Q_1 (\tfrac{i}{2})}{Q_0 (\tfrac{i}{2})}
\, , \nonumber\\
\varepsilon^+
&=&
i  \frac{Q_1' (\tfrac{i}{2})}{Q_0 (\tfrac{i}{2})}
\, , \nonumber
\ea
and analogous expressions whose argument has reverse sign being related to the above
via
\ba
\alpha^- &=& (\alpha^+)^\ast = \alpha^+ \equiv \alpha
\, , \qquad
\beta^- = - (\beta^+)^\ast = - \beta^+ = - \beta
\, , \nonumber\\
\chi^- &=& (\chi^+)^\ast  = \chi^+  \equiv \chi
\, , \qquad
\delta^- = (\delta^+)^\ast = \delta^+ \equiv \delta
\, , \nonumber\\
&&\qquad\qquad
\varepsilon^-
=
- (\varepsilon^+)^\ast = - \varepsilon^+ = - \varepsilon
\, .
\ea
They can be written in terms of harmonic sums making use of the explicit solution to
the Baxter
equation and read
\ba
\alpha
&=&
- S_1^2 + S_{-2} \, , \\
\beta
&=&
- \tfrac{1}{24}(S_1^3-S_3+3S_{1,-2}-3S_{-2,1}) \, , \nonumber\\
\chi
&=&
\ft{1}{2}
(S_{-4}-S_{4}-S_{-3,1}+S_{-2,-2}-S_{-2,2}-S_{1,-3}+S_{1,3}-S_{2,-
2}+S_{2,2}+S_{3,1})
\nonumber\\
&&+
S_{-2,1,1}+S_{1,1,-2}-S_{1,1,2}-S_{1,2,1}-S_{2,1,1}+2S_{1,1,1,1}
\nonumber\\
&&+
\ft{1}{4} \gamma_0 (S_{-3}-S_{3}-S_{-2,1}+S_{1,-2}+S_{1,2}+S_{2,1})
\, , \nonumber\\
\delta &=& a_1
\, , \nonumber\\
\varepsilon
&=&
- \gamma_0 (S_{-2}+S_{2}) + 2(S_{-3}-S_{3}-2S_{-2,1}) + 2 a_1 S_{1}
\, , \nonumber
\ea
expressed in terms of nested harmonic sums
\be
\label{NestHarmSum}
S_{a_1, \dots, a_k} \equiv S_{a_1, \dots, a_k} (M)
=
\sum_{\ell_1 = 0}^M \frac{(-1)^{{\rm sign} (a_1)}}{\ell_1^{a_1}}
\sum_{\ell_2 = 0}^{\ell_1} \frac{(-1)^{{\rm sign} (a_2)}}{\ell_2^{a_2}}
\dots
\sum_{\ell_k = 0}^{\ell_{k - 1}} \frac{(-1)^{{\rm sign} (a_k)}}{\ell_k^{a_k}}
\ee
%%%%%%%%%%%%%%%%%%%%%%%%%%%%%%%%%%%%%%%%%%%%%%%%%%%%%%%%%%%%%%%
\section{Continuous Hahn and Wilson polynomials}
\label{WilsonPolynomials}
%%%%%%%%%%%%%%%%%%%%%%%%%%%%%%%%%%%%%%%%%%%%%%%%%%%%%%%%%%%%%%%

Continuous Hahn polynomials are defined as~\cite{AskWil85}
\be
p_n(u, a, b, c, d) = i^n\,\frac{(a+c)_n\,(a+d)_n}{n!}\,{}_3F_2\left(
\left.
\begin{array}{c}
-n,\ n+a+b+c+d-1,\ a+i\,u\\
a+c,\ a+d
\end{array}
\right| 1\right).
\ee
They obey the functional relation ($P(u) \equiv p_n(\cdots)$)
\ba
\lefteqn{(c-i\, u) (d-i\, u)\, P(u+i) + (a+i\, u) (b+i\, u)\, P(u-i) + } &&  \\
&& (2\, u^2 -i \,(a+b-c-d)\, u -n^2+(-a-b-c-d+1)\, n-a \,b-c \,d) \,P(u) =
0 \,,\nonumber
\ea
or
\ba
(c-i\, u) (d-i\, u)\, [P(u+i)-P(u)] &+& (a+i\, u) (b+i\, u)\, [P(u-i)-P(u)] +
\\
&&  \hskip -3cm-(n^2+n\,(a+b+c+d-1)\,P(u) = 0 \, .\nonumber
\ea

Wilson polynomials are defined as
\begin{eqnarray}
W_n(u^2, a, b, c, d) &=& (a + b)_n (a + c)_n (a + d)_n \nonumber\\
&&\times{}_4 F_3
\left(
\left.
\begin{array}{c}
- n,\  n + a+b+c+d-1, \ a + i\ u,\ a - i\ u \\
a + b ,\ a + c ,\ a + d
\end{array}
\right| 1 \right).
\end{eqnarray}
They obey the functional relation ($P(u) \equiv W_n(u^2, \cdots)$)
\ba
\lefteqn{(c-i u) (i a+u) (i b+u) (i d+u) (2 u-i) P(u+i) } && \\
&& -(c+i u) (u-i a) (u-i b) (u-i d) (2 u+i) P(u-i) + 2\,i\,u\,A\,P(u) =
0 \,,\nonumber
\ea
with
\ba
A &=& 2\,u^4 + u^2\,(-4 n^2-4 (a+b+c+d-1) n+a + b + c + d  + \nonumber\\
&& \qquad\qquad -2 a b-2 a c-2 b c-2 a d-2 b d-2 c d) + \nonumber\\
&& -n^2+(-a-b-c-d+1) n-a b c-a b d-a c d+2 a b c d-b c d \,.
\ea
It can also be rewritten as
\begin{eqnarray}
\label{WilsonEq}
n (n + s - 1) Y (u) = B (u) \left[ Y (u + i) - Y (u) \right] + B (- u) \left[ Y
(u - i) - Y (u) \right]
\, ,
\end{eqnarray}
with
\be
B (u) = - \frac{(u + i a)(u + i b)(u + i c)(u + i d)}{2 u (2u + i)}
\, .
\ee
The relation
\be
p_M\left(u, \frac{1}{2}, \frac{1}{2}, \frac{1}{2}, \frac{1}{2}\right) =
\mbox{constant}\times W_{M/2}\left(u,  \frac{1}{2}, \frac{1}{2}, \frac{1}{2},
0\right),
\ee
is immediately proved since the two recurrence relation are equal for the above
choice of parameters. Looking at the normalization, one can check that the
constant is
indeed equal to one.

The comparison of the two-loop Baxter equation and equations obeyed by
polynomials yield the
identifications for the continuous Hahn
\be
n = M \, , \qquad a = c = \ft12 + \ft{i}{\sqrt{2}} g + \ft14 g^2 \gamma_0 \, ,
\qquad\qquad b = d = \ft12 - \ft{i}{\sqrt{2}} g + \ft14 g^2 \gamma_0
\, ,
\ee
and Wilson polynomials,
\begin{eqnarray}
n &=& \ft12 M \, , \qquad s = \ft32 + \ft12 g^2 \gamma_0 \, ,
\qquad a = \ft12 \, , \qquad d = 0
\, , \\
b &=& \ft12 + \ft{i}{\sqrt{2}} g + \ft14 g^2 \gamma_0 \, ,
\qquad\qquad c = \ft12 - \ft{i}{\sqrt{2}} g + \ft14 g^2 \gamma_0
\, ,
\end{eqnarray}
respectively.

%%%%%%%%%%%%%%%%%%%%%%%%%%%%%%%%%%%%%%%%%%%%%%%%%%%%%%%%%%%%%%%%%%%%5%
\section{Derivatives of the Baxter functions}
\label{App:Derivatives}
%%%%%%%%%%%%%%%%%%%%%%%%%%%%%%%%%%%%%%%%%%%%%%%%%%%%%%%%%%%%%%%

The $n$-th derivative of $Q^{(0)}(u)$ evaluated at $u=i/2$ is a combination of
harmonic sums with uniform transcendentality $n$
and multi-indices containing only $1$ and $\pm 2$. The first cases are
\ba
Q_0^\prime (\ft{i}{2}) &=& -2\, i\, \,S_1, \\
Q_0^{\prime\prime} (\ft{i}{2}) &=& -4\, \left(S_{-2}-S_2+2 S_{1,1}\right), \\
Q_0^{\prime\prime\prime} (\ft{i}{2}) &=& -24\, i\, \left(S_{-2,1}-S_{1,-
2}+S_{1,2}+S_{2,1}-2 S_{1,1,1}\right), \\
Q_0^{(4)} (\ft{i}{2}) &=& 96\, (S_{-2,-2}-S_{-2,2}-S_{2,-2}+S_{2,2}+2 S_{-
2,1,1}-2 S_{1,-2,1}+2 S_{1,1,-2}+ \nonumber\\
&& -2 S_{1,1,2} -2 S_{1,2,1}-2 S_{2,1,1}+4 S_{1,1,1,1}), \\
Q_0^{(5)} (\ft{i}{2})  &=& -960\, i\, (S_{-2,-2,1}-S_{-2,1,-2}+S_{-2,1,2}+S_{-
2,2,1}+S_{1,-2,-2}-S_{1,-2,2} + \nonumber\\
&& -S_{1,2,-2}+S_{1,2,2}+S_{2,-2,1}-S_{2,1,-2}+S_{2,1,2}+S_{2,2,1}-2 S_{-
2,1,1,1}+ \nonumber\\
&& 2 S_{1,-2,1,1} -2 S_{1,1,-2,1}+2 S_{1,1,1,-2}-2
   S_{1,1,1,2}-2 S_{1,1,2,1}-2 S_{1,2,1,1} + \nonumber\\
&& -2 S_{2,1,1,1}+4 S_{1,1,1,1,1}), \\
Q_0^{(6)} (\ft{i}{2})  &=& -5760\, \,(S_{-2,-2,-2} -S_{-2,-2,2}-S_{-2,2,-2}+S_{-
2,2,2}-S_{2,-2,-2}+\nonumber\\
&& + S_{2,-2,2}+S_{2,2,-2}-S_{2,2,2}+2 S_{-2,-2,1,1}-2 S_{-2,1,-2,1}+2 S_{-
2,1,1,-2}+\nonumber\\
&& -2 S_{-2,1,1,2}-2 S_{-2,1,2,1}-2 S_{-2,2,1,1}+2 S_{1,-2,-2,1}-2
   S_{1,-2,1,-2}+\nonumber\\
&& + 2 S_{1,-2,1,2}+2 S_{1,-2,2,1}+2 S_{1,1,-2,-2}-2 S_{1,1,-2,2}-2 S_{1,1,2,-
2}+2 S_{1,1,2,2}+ \nonumber\\
&& +2 S_{1,2,-2,1} -2 S_{1,2,1,-2}+2 S_{1,2,1,2}+2 S_{1,2,2,1}-2 S_{2,-2,1,1}+2
S_{2,1,-2,1} + \nonumber \\
&& -2 S_{2,1,1,-2}+2
   S_{2,1,1,2}++ 2 S_{2,1,2,1}+2 S_{2,2,1,1}+4 S_{-2,1,1,1,1}-4 S_{1,-2,1,1,1}+
\nonumber\\
&& 4 S_{1,1,-2,1,1} -4 S_{1,1,1,-2,1}+4 S_{1,1,1,1,-2}-4 S_{1,1,1,1,2}-4
S_{1,1,1,2,1}-4 S_{1,1,2,1,1} + \nonumber\\
&& -4 S_{1,2,1,1,1}  -4 S_{2,1,1,1,1}+8 S_{1,1,1,1,1,1}).
\ea
Let us write the two-loop Baxter polynomial as
\be
Q_1(u) = a_1(M)\,Q_0(M) + \delta Q_1 (u),
\ee
where $a_1$ is the same as in Eq.~(\ref{Q1}).
The derivatives of $Q^{(1)}(u)$ are expressed in terms of the derivatives of
$Q^{(0)}(u)$ and those of $\delta Q^{(1)}(u)$.
The $n$-th derivative of $\delta Q^{(1)}(u)$ can be expressed in terms of
harmonic sums with uniform transcendentality $n+2$. The first cases are
\begin{eqnarray}
\delta Q_1 (\ft{i}{2})  &=& 0, \\
\delta Q_1^{\prime} (\ft{i}{2})  &=& -2\, i\, \left(2 S_{-3}-3 S_{-2,1}-S_{1,-
2}-S_{1,2}-S_{2,1}\right), \\
\delta Q_1^{\prime\prime} (\ft{i}{2})  &=& 8\, (S_{-3,1}+S_{-2,2}-S_{1,-3}-
S_{1,3}-S_{2,-2}-2 S_{2,2}-S_{3,1} +\\
&& -2 S_{-2,1,1}+2 S_{1,-2,1}+2 S_{1,1,-2}+2 S_{1,1,2}+2 S_{1,2,1}+2 S_{2,1,1}),
\nonumber \\
\delta Q_1^{\prime\prime\prime} (\ft{i}{2})  &=& -24\, i\,(2 S_{-4,1}-S_{-3,-
2}+5 S_{-3,2}+S_{-2,-3}+S_{-2,3}-S_{2,-3}+\\
&& + 3 S_{2,3}+S_{3,-2}+3 S_{3,2}+2 S_{4,1}-8 S_{-3,1,1}-4 S_{-2,-2,1}+2 S_{-
2,1,-2}+\nonumber\\
&& -6 S_{-2,1,2}-6 S_{-2,2,1}-4 S_{1,1,3}-2 S_{1,2,-2}-6 S_{1,2,2}-4
   S_{1,3,1}-2 S_{2,1,-2}+\nonumber\\
&& -6 S_{2,1,2}-6 S_{2,2,1}-4 S_{3,1,1}+10 S_{-2,1,1,1}-2 S_{1,-2,1,1}+2
S_{1,1,-2,1}+\nonumber\\
&& + 6 S_{1,1,1,-2}+6 S_{1,1,1,2}+6 S_{1,1,2,1}+6 S_{1,2,1,1}+6 S_{2,1,1,1}),
\nonumber \\
\delta Q_1^{(4)} (\ft{i}{2}) &=& 384\,(S_{-3,-3}-S_{-3,3}-S_{3,-3}+S_{3,3}-S_{-
3,-2,1}+2 S_{-3,1,2}+ \\
&& + 2 S_{-3,2,1}+S_{-2,1,3}+2 S_{-2,2,2}+S_{-2,3,1}-S_{1,-4,1}-2 S_{1,-
3,2}+\nonumber\\
&& -S_{1,-2,-3}+S_{1,2,-3}-2 S_{1,2,3}-2
   S_{1,3,2}-S_{1,4,1}-S_{2,-3,1}-S_{2,-2,2}+\nonumber\\
&& + S_{2,1,-3}-2 S_{2,1,3}-3 S_{2,2,2}-2 S_{2,3,1}+S_{3,-2,1}-2 S_{3,1,2}-2
S_{3,2,1}+\nonumber\\
&& -3 S_{-3,1,1,1}-S_{-2,1,-2,1}-3 S_{-2,1,1,2}-3 S_{-2,1,2,1}-3 S_{-2,2,1,1}+3
S_{1,-3,1,1}+\nonumber\\
&& + 2
   S_{1,-2,-2,1}-S_{1,-2,1,-2}+2 S_{1,-2,1,2}+2 S_{1,-2,2,1}+S_{1,1,-
3,1}+S_{1,1,-2,2}+\nonumber\\
&& -S_{1,1,1,-3}+3 S_{1,1,1,3}+S_{1,1,2,-2}+4 S_{1,1,2,2}+3
S_{1,1,3,1}+S_{1,2,1,-2}+\nonumber\\
&& + 4 S_{1,2,1,2}+4 S_{1,2,2,1}+3
   S_{1,3,1,1}+S_{2,-2,1,1}+S_{2,1,1,-2}+4 S_{2,1,1,2}+\nonumber\\
&& + 4 S_{2,1,2,1}+4 S_{2,2,1,1}+3 S_{3,1,1,1}+4 S_{-2,1,1,1,1}-4 S_{1,-
2,1,1,1}+\nonumber\\
&& -4 S_{1,1,1,1,-2}-4 S_{1,1,1,1,2}-4 S_{1,1,1,2,1}-4 S_{1,1,2,1,1}-4
S_{1,2,1,1,1}-4
   S_{2,1,1,1,1}).\nonumber
\end{eqnarray}
%
%%%%%%%%%%%%%%%%%%%%%%%%%%%%%%%%%%%%%%%%%%%%%%%%%%%%%%%%%%%%%%%%%%%%%%%%%%%%
\section{Mellin transformation}
\label{MellinTransformation}
%%%%%%%%%%%%%%%%%%%%%%%%%%%%%%%%%%%%%%%%%%%%%%%%%%%%%%%%%%%%%%%%%%%%%%%%%%%%

In this appendix we devise a very efficient formalism for finding solutions to
the second-order
finite difference Baxter equation focusing on non-polynomial inhomogeneities.
The results that
we will present are very general, covering all possible powers of non-polynomial
contributions,
$(u^\pm)^{- k} Q_\ell (u)$, and are therefore applicable to any order of
perturbation theory.
The restriction and application of the machinery to three- and four-loop non-
polynomial
inhomogeneities, which are the main objective of the current study, are given in
sections
\ref{npThreeloop} and \ref{npFourloop}, respectively. We also employ it to the
dressing-induced
five-loop non-polynomial part in section \ref{five-loop dresci}. Below, we first
introduce the
Mellin transform for the Baxter function and accompanying (inverse) polynomial
dressing factors
in the spectral parameter. Then we perform a variable transformation which is
particularly
convenient for carrying out Mellin convolutions and inverse transform.

To start with let us introduce the Mellin transform of the main ingredients. For
the Baxter
function $Q (u)$, where the loop-order subscript is dropped being irrelevant, it
takes the
form \cite{FadKor95}
\ba
Q(u) = K \int_0^{\infty} d\omega \, {\omega}^{iu-1} Q(-\omega)\,, \quad
K = \frac{1}{\Gamma(iu)\Gamma(1-iu)}
\, .
\label{M1}
\ea
While for the function with the argument shifted by $i$, i.e., $Q(u \pm i)$, we
find
that in Mellin space this yields a multiplication by a power of $\omega$
\ba
Q(u\pm i) = -K \int_0^{\infty} d\omega \, {\omega}^{iu-1}
\left\{{\omega}^{\mp1} Q(-\omega)\right\}
= K \int_0^{\infty} d\omega \, {\omega}^{iu-1}
\left\{-{\omega}^{\mp1} Q(-\omega)\right\} \, .
\label{M2}
\ea
Next, turning to terms where $Q(u)$ is accompanied by a positive power of a
polynomial
in the spectral parameter of the form $(u+bi)^L \equiv \lambda^L$, we get at
first
\ba
(u+bi)^L \, Q(u) = \lambda^L \, K \int_0^{\infty} d\omega \,
{\omega}^{i\lambda-1} {\omega}^b Q(-\omega) \, .
\label{M3}
\ea
Then, we can re-express the product $\lambda^L {\omega}^{i\lambda-1}$ by a
differential
operator acting on the exponent, i.e., $- i (d/d\omega \, \omega)^L {\omega}^{i
\lambda-1}$,
and subsequently integrate by parts to find
\be
(u+bi)^L \, Q(u)
=
K \int_0^{\infty} d\omega \,
{\omega}^{iu-1} \, {\omega}^{-b} \,
{\left[  i\omega \frac{d}{d\omega} \right]}^L
\left\{ {\omega}^b Q(-\omega)\right\}
\, .
\label{M4}
\ee
Similarly, for the Baxter function with the shifted argument $Q(u\pm i)$, we get
the
expression
\ba
(u+bi)^L \, Q(u\pm i)
&=&  K \int_0^{\infty} d\omega \,
{\omega}^{iu-1} \, {\omega}^{-b} \,
{\left[  i\omega \frac{d}{d\omega} \right]}^L
\left\{ -{\omega}^{b\mp 1} Q(-\omega)\right\}
\, .
\label{M5}
\ea

In the same vein we can consider terms with inverse powers of the same
polynomial, i.e.,
$(u+bi)^{- L}$. First, notice that for $L=1$, its Mellin integral representation
\begin{equation}
\frac{1}{u+bi}
= \pm i \int_0^{\infty} d\omega \, {\omega}^{iu-1-b} \theta(\pm 1 \mp \omega)
=
\left\{
\begin{array}{ll}
+ i \int_0^{1} d\omega \, {\omega}^{iu-1-b}
\\[2mm]
- i \int_1^\infty d\omega \, {\omega}^{iu-1-b}
\end{array}
\right.
\, , \qquad \mbox{for} \qquad b \lessgtr 0
\, ,
\label{M6}
\end{equation}
is given in terms of the the Heaviside step function $\theta(x)$, defined
conventionally as
\begin{equation}
\theta(x) =
\left\{
\begin{array}{ll}
1 \, , & \quad \mbox{for} \quad x \geq 0 \,; \\
0 \, , & \quad \mbox{for} \quad x < 0 \,.
\end{array}
\right.
\label{M7}
\end{equation}
The other observation is the well-known fact that the Mellin transform of a
product of functions
is realized as a convolution of their Mellin transforms. Namely, if a function
$M_j (u)$ is expressed
in terms of its Mellin transform
\begin{equation}
M_j (u) \, = \, \int_0^{\infty} d\omega \, {\omega}^{iu-1} F_j (\omega) \,,
\label{M8}
\end{equation}
then the product of $M_1 (u) M_2(u)$ corresponds to
\begin{equation}
M_1(u)\, M_2(u) \,
=
\, \int_0^{\infty} d\omega \, {\omega}^{iu-1}
\, \int_0^{\infty} \frac{d\omega_1}{\omega_1} \, F_1(\omega_1) \,
F_2\left(\frac{\omega}{\omega_1}\right) \, ,
\label{M9}
\end{equation}
which we will write formally as
\be
M_1(u)\, M_2(u)
\, \stackrel{M^{-1}}{\to} \,
\int_0^{\infty} \frac{d\omega_1}{\omega_1} \, F_1(\omega_1) \,
F_2\left(\frac{\omega}{\omega_1}\right)
\, ,
\label{M10}
\ee
where the symbol $\stackrel{M^{-1}}{\to}$ implies the inverse Mellin transform.
Putting the
two results together, leads to
\ba
\frac{1}{u+bi} \, Q(u)  \, \stackrel{M^{-1}}{\to} \,
&\pm& i K \, \int_0^{\infty} \frac{d\omega_1}{\omega_1} \, Q(- \omega_1) \,
\left(\frac{\omega}{\omega_1}\right)^{-b}  \,
\theta \left(\pm 1 \mp \ft{\omega}{\omega_1} \right)
\nonumber\\
&=&
\left\{
\begin{array}{l}
+ i K \, \int_{\omega}^{\infty}  \frac{d\omega_1}{\omega_1} \,
\left(\frac{\omega}{\omega_1}\right)^{-b}  \, Q(-\omega_1)
\\[2mm]
- i K \, \int_0^{\omega}  \frac{d\omega_1}{\omega_1} \,
\left(\frac{\omega}{\omega_1}\right)^{-b}  \, Q(-\omega_1)
\end{array}
\right.
\, , \qquad \mbox{for} \qquad b \lessgtr 0
\, .
\label{M11}
\ea
Now, upon differentiating both sides of this equation w.r.t.\ $b$, we obtain the
final result
\be
\frac{L!}{(u+bi)^L} \, Q(u)  \, \stackrel{M^{-1}}{\to} \,
\left\{
\begin{array}{l}
-(-i)^{L+1} K \, \int_{\omega}^{\infty}  \frac{d\omega_1}{\omega_1} \,
{\left(\ln \frac{\omega}{\omega_1}\right)}^{L}  \,
{\left(\frac{\omega}{\omega_1}\right)}^{-b}  \, Q(-\omega_1)
\\[2mm]
+ (-i)^{L+1} K \, \int_0^{\omega}  \frac{d\omega_1}{\omega_1} \,
{\left(\ln \frac{\omega}{\omega_1}\right)}^{L}  \,
{\left(\frac{\omega}{\omega_1}\right)}^{-b}  \, Q(-\omega_1)
\end{array}
\right.
\, , \qquad \mbox{for} \qquad b \lessgtr 0
\, .
\label{M12}
\ee
And for the shifted Baxter polynomial $Q(u \pm i)$, this is replaced by
\ba
&&
\frac{L!}{(u+bi)^L} \, Q(u\pm i)
\\
&&\qquad
\stackrel{M^{-1}}{\to} \,
\left\{
\begin{array}{l}
-(-i)^{L+1} K \, \int_{\omega}^{\infty}  \frac{d\omega_1}{\omega_1} \,
{\left(\ln \frac{\omega}{\omega_1}\right)}^{L}  \,
{\left(\frac{\omega}{\omega_1}\right)}^{-b}  \,
\left[ -{\omega_1}^{\mp 1} Q (-\omega_1) \right]
\\[2mm]
+ (-i)^{L+1} K \, \int_0^{\omega} \, \frac{d\omega_1}{\omega_1} \,
{\left(\ln \frac{\omega}{\omega_1}\right)}^{L}  \,
{\left(\frac{\omega}{\omega_1}\right)}^{-b}  \,
\left[ -{\omega_1}^{\mp 1} Q (-\omega_1) \right]
\end{array}
\right.
\, , \qquad \mbox{for} \qquad b \lessgtr 0
\, . \nonumber
\ea
%
%%%%%%%%%%%%%%%%%%%%%%%%%%%%%%%%%%%%%%%%%%%%%%%%%%%%%%%%%%%%%%%%%%%%%%%%%%%%%%%
\subsection{Change of variables}
%%%%%%%%%%%%%%%%%%%%%%%%%%%%%%%%%%%%%%%%%%%%%%%%%%%%%%%%%%%%%%%%%%%%%%%%%%%%%%%

To proceed further, we introduce the variable $z$ as in \cite{FadKor95},
\be
\omega \, = \, \frac{z}{1-z} \,,
\label{M13}
\ee
such that the Mellin transform (\ref{M1}) takes the form
\be
Q(u) \, = \, K_1 \int^1_0 dz \, z^{iu-1/2} \, (1-z)^{-iu-1/2} \, \Psi(z) \,,
\label{M14}
\ee
where we introduced a new function (see Ref.\ \cite{KotRejZie08})
\be
Q(z) \, = \, \sqrt{z(1-z)} \, \Psi(z)
\, ,
\label{M15}
\ee
with the corresponding normalization factor
\be
K_1 \, = \, \frac{1}{\Gamma(\ft12 - iu)\Gamma(\ft12 + iu)}
\, .
\label{M17}
\ee
The latter leads to the unit normalization factor in Eq.\ (\ref{Q03F2}).

Absorbing all terms of the all-order Baxter equation beyond one loop into inhomogeneities
$\widehat{Q}(u)$ on the right-hand side, we can write it as
\be
(u^+)^2 Q(u + i) + (u^-) ^2 Q(u - i) - t_0 (u) Q(u)
=
\widehat{Q}(u)
\, .
\label{M18}
\ee
While the corresponding equation in Mellin space reads (we omitted the overall factor $K_1 \sqrt{z(1-z)}$
on both left- and right-hand sides)
\be
z(1 - z) \Psi''(z) + (1 - 2z) \Psi'(z) + M(M + 1) \Psi(z)
= \widehat{\Psi}(z)
\, ,
\label{M19}
\ee
and where the primes on $\Psi (z)$, e.g., ~$\Psi'(z)$, correspond to the differentiation w.r.t.\ the
variable $z$.

The functions $Q(u)$ and $\Psi(z)$ admit perturbative expansions
\be
Q(u) \, = \, \sum_{\ell = 0}^{\infty} \, g^{2 \ell} \, Q_\ell (u)
\, , \qquad
\Psi(z) \, = \, \sum_{\ell = 0}^{\infty} \, g^{2 \ell} \, \Psi_\ell (z) \,,
\label{M16b}
\ee
with their coefficients taking the form
\ba
Q_\ell (u) \, = \,
\sum_{k=0}^{M} \, R_{\ell, k}(M) \, P_k(u)
\, , \qquad
\Psi_\ell (z)   \, = \,
\sum_{k=0}^{M} \, R_{\ell, k}(M) \, z^k
\, .
\label{M16c}
\ea
In the following, we will reserve the convention $R_{0, k}(M) \, = \, R_{k}(M)$ for the leading
order term, with the functions $R_k(M)$ and $P_k(u)$ introduced in Eq.\ (\ref{N1}).

Note that the solution of (\ref{M19}) can be written as a sum of the $z^k$- and $(1-z)^k$-expansions,
i.e.,
\be
\Psi_{\ell} (z)   \, = \,
\frac{1}{2} \, \sum_{k=0}^{M} \, \Bigl[ z^k + (1-z)^k \Bigr] \,
\, R_{\ell,k}(M) \, ,
\label{M16ca}
\ee
which corresponds to the real part of its Mellin transform counterpart (\ref{M16c}), i.e.,
\be
Q_{\ell} (u) \, = \,
\sum_{k=0}^{M} \Re{\rm e}[ P_{k}(u)] \,
\, R_{\ell,k}(M)
\, ,
\label{M16cb}
\ee
and is the solution that is sought for.

%%%%%%%%%%%%%%%%%%%%%%%%%%%%%%%%%%%%%%%%%%%%%%%%%%%%%%%%%%%%%%%%%%%%%%%
\subsection{General properties}
%%%%%%%%%%%%%%%%%%%%%%%%%%%%%%%%%%%%%%%%%%%%%%%%%%%%%%%%%%%%%%%%%%%%%%%

Consider the polynomial solution (\ref{M16c}) to the Baxter equation (\ref{M19}). 
The left-hand side of the latter can be cast in the form
\be
\sum_{k=0}^{M-1} \, \biggl[ \, (k+1)^2 R_{\ell, k+1}(M) -
(M-k)(k+1+M) R_{\ell, k}(M) \, \biggr]
\, z^k  \, ,
\label{M20}
\ee
and contains the maximal $(M-1)$-th power of the $z$-variable compared to the solution (\ref{M16c})
which is a $M$-th order polynomial.

It is convenient to introduce the new coefficients $r_{\ell, k+1}(M)$ as
\be
r_{\ell, k}(M) = \frac{R_{\ell, k}(M)}{R_{k}(M)}
\, ,
\label{M21}
\ee
with $R_k(M)$ given in Eq.\ (\ref{N1}). Then, since it obeys the recurrence relation
\be
R_{k}(M) \, = \,  R_{k+1}(M) \frac{(k+1)^2}{(M-k)(k+1+M)} \, ,
\label{M22ac}
\ee
we can replace (\ref{M20}) by
\be
\sum_{k=0}^{M-1} \, z^k  \,  R_{k}(M)  \, (k+1)^2
\biggl[ r_{\ell, k+1}(M) - r_{\ell, k}(M) \, \biggr] \, .
\label{M23a}
\ee
At leading order of perturbation theory, the right-hand side $\widehat{\Psi}(z)$ of Eq.\ (\ref{M19})
vanishes, and the equation for $r_{\ell, k}$ is simply
\be
r_{\ell, k+1}(M) - r_{\ell, k}(M) = 0
\, .
\label{M24}
\ee
The solution is
\be
r_{\ell, k}(M) = r_{\ell, 0}(M) \, , \qquad
R_{0, k}(M) \, = r_{\ell, 0}(M) \cdot \, R_{k}(M) \, ,
\label{M25}
\ee
where $r_{\ell, 0}(M)$ is a constant. The possible presence of nontrivial contributions from the
$(1-z)^k$ part of the expansion in the right-hand side of Eq.\ (\ref{M19}) does not alter any of the
above results because its left-hand side is symmetric under the exchange $z \to (1-z)$.

Note that in order to have the solution of Eq.\ (\ref{M19}) in the form of (\ref{M16c}), the inhomogeneities
$\widehat{Q}(u)$ and $\widehat{\Psi}(z)$ should be expanded in the same basis of functions
\ba
\widehat{Q}(u) \, = \, \sum_{k=0}^{M-1} \, P_k(u) \, B_{k+1}
\, , \qquad
\widehat{\Psi}(z)  \, = \,
\sum_{k=0}^{M-1} \, B_{k+1}
\,  z^k ,
\label{M26a}
\ea
and, respectively,
\ba
\widehat{Q}(u) \, = \, \sum_{k=0}^{M-1} \, \Re{\rm e}[ P_{k}(u)] \, B_{k+1}
\, , \qquad
\widehat{\Psi}(z)  \, = \,
\frac{1}{2} \, \sum_{k=0}^{M} \, \Bigl[ z^k + (1-z)^k \Bigr] \,
B_{k+1} \, ,
\label{M26b}
\ea
if the $(1-z)^k$-expansion constitutes a nontrivial contribution.

Going beyond leading order of perturbation theory, the inhomogeneities in the Baxter equation
affect the recurrence relation for the expansion coefficients (\ref{M21}) as follows
\be
r_{\ell, k+1}(M) - r_{\ell, k}(M) =  \frac{B_{k+1}}{R_{k+1}(M)(k+1)^2} \,,
\label{M27}
\ee
such that the iterative solution to it reads
\be
r_{\ell, k}(M)
\, = \, r_{\ell, 0}(M)
+
\sum_{l=1}^{k} \frac{B_{l}}{R_l(M)l^2}
\, .
\label{M28a}
\ee
Here, the first term in (\ref{M28a}) corresponds to the general solution of the homogeneous
equation and the second one is a particular solution of the inhomogeneous equation (\ref{M19}).

%%%%%%%%%%%%%%%%%%%%%%%%%%%%%%%%%%%%%%%%%%%%%%%%%%%%%%%%%%%%%%%%%%%%%%%
\subsection{Polynomial inhomogeneities}
%%%%%%%%%%%%%%%%%%%%%%%%%%%%%%%%%%%%%%%%%%%%%%%%%%%%%%%%%%%%%%%%%%%%%%%

Starting from two-loop order, the right-hand side of Eq.\ (\ref{M18}) is nonzero and
contains power-series contribution in $\sim (u^\pm)^k$ with $k\leq 1$. The terms $\sim u^1$
and $\sim u^0$ induce polynomial contributions. The bulk of them can be calculated directly
in the $u$-space (see Section 4). However, it is instructive to consider their calculation
also in the Mellin transform $z$-space, because, starting from $4$-loop, nonpolynomial effects
from lower orders of perturbation theory will re-emerge through these terms as well.

For simplicity and since this is all one needs for the present analysis, we limit our
consideration to contributions at $n$-th order of perturbation theory from Baxter functions
of one loop order lower. All other cases can be treated similarly since the only 
difference will be in the coefficients accompanying $u^1$ and $u^0$.

%%%%%%%%%%%%%%%%%%%%%%%%%%%%%%%%%%%%%%%%%%%%%%%%%%%%%%%%%%%%%%%%%%%%%%%
\subsubsection{Contributions $\sim u^1$}
%%%%%%%%%%%%%%%%%%%%%%%%%%%%%%%%%%%%%%%%%%%%%%%%%%%%%%%%%%%%%%%%%%%%%%%

For the case at hand, the corresponding inhomogeneity in the right-hand side has the 
following form (see also Eq. (5.48))
\be
iu^+ \widehat{Q}_1(u + i) - iu^- \widehat{Q}_1(u - i) +(2M+1) \widehat{Q}_1(u) \,,
\label{MA1}
\ee
with a function $\widehat{Q}_1(u)$. After the Mellin transformation to $z$-space (similar to
Eq.\ (\ref{M14})) and cancelling the overall coefficient $K_1 \, \sqrt{z(1-z)}$, we find
\be
(1 - 2z) \widehat{\Psi}_1'(z) + 2M \widehat{\Psi}_1(z)\, .
\label{MA2}
\ee
Let now $\widehat{Q}_1(u)$ and $\widehat{\Psi}_1(z)$ have the following expansions (analogous
to (\ref{M16ca}) and (\ref{M16cb}))
\be
\widehat{Q}_1 (u) \, = \,
\sum_{k=0}^{M} \Re{\rm e}[ P_{k}(u)] \,
\, \widehat{F}_1(k)
\, , \qquad
\widehat{\Psi}_1 (z)   \, = \,
\frac{1}{2} \, \sum_{k=0}^{M} \, \Bigl[ z^k + (1-z)^k \Bigr] \,
\, \widehat{F}_1(k) \,.
\label{MA3}
\ee
Then, the corresponding particular solutions, $\overline{Q}_1 (u)$ and 
$\overline{\Psi}_1 (z)$, of the Baxter equation and
its Mellin analogue, can be found in terms of the expansion coefficients $\widehat{F}_1 (k)$
\be
\overline{Q}_1 (u) \, = \,
\sum_{k=0}^{M} \Re{\rm e}[ P_{k}(u)] \,
\, F_1(k)
\, , \qquad
\overline{\Psi}_1 (z)   \, = \,
\frac{1}{2} \, \sum_{k=0}^{M} \, \Bigl[ z^k + (1-z)^k \Bigr] \,
\, F_1(k) \, ,
\label{MA4}
\ee
where the coefficients $F_1(k)$ are determined by $\widehat{F}_1(k)$ by Eqs.\ (\ref{M27}) and
(\ref{M28a}).

It is possible to consider the expansions in $z^k$ and $(1-z)^k$ independently.
Addressing the $z^k$-expansion first, we have from (\ref{MA2}) and (\ref{MA3})
\ba
&&\sum_{k=0}^{M-1} \, z^k \,
\Bigl[ (k+1) \widehat{F}_1(k+1) + 2(M-k) \widehat{F}_1(k) \Bigr] \nonumber \\
&& =
\sum_{k=0}^{M-1} \, z^k \, R_{k+1}(M) \, (k+1)^2 \,
\left[ \frac{\widehat{f}_1(k+1)}{k+1} - \frac{2\widehat{f}_1(k)}{k+1+M}
\right] \, ,
\label{MA5}
\ea
where we factored out the $R_{k}$ dependence and as a result introduced new coefficients
\be
F_1(k) = R_{k}(M)f_1(k)
\, , \qquad
\widehat{F}_1(k) = R_{k}(M)\widehat{f}_1(k)
\, .
\label{MA6}
\ee
Next, putting Eqs.\ (\ref{M26a}), (\ref{M27}) and (\ref{MA5}) together, we get for $f_1(k)$
the following representation
\be
f_1(k)= f_1(0) + \sum_{m=1}^k \, \left[ \frac{\widehat{f}_1(m)}{m} -
\frac{2\widehat{f}_1(m-1)}{m+M} \right] \, .
\label{MA7}
\ee
Finally, when we set $\widehat{Q}_1(u) = Q_0 (u)$, i.e.~to the leading order Baxter polynomial,
the coefficients can be found explicitly and read
\be
\widehat{f}_1(k)= 1
\, , \qquad
f_1(k) = S_1(k) +2 \Bigl(S_1(k+M)-S_1(M) \Bigr)
\, .
\label{MA8}
\ee
In this way we thus reproduce the $\gamma_0$-part of the $Q_1$ Baxter function in (\ref{NN2+}).

%%%%%%%%%%%%%%%%%%%%%%%%%%%%%%%%%%%%%%%%%%%%%%%%%%%%%%%%%%%%%%%%%%%%%%%
\subsubsection{Contributions $\sim u^0$}
%%%%%%%%%%%%%%%%%%%%%%%%%%%%%%%%%%%%%%%%%%%%%%%%%%%%%%%%%%%%%%%%%%%%%%%

For the $u^0$ term, the corresponding right-hand side has the following form (see also Eq.\ (5.48))
\be
\widehat{Q}_2(u + i) + \widehat{Q}_2(u - i) - 2 \widehat{Q}_2(u)
\, ,
\label{MB1}
\ee
in terms of a function $\widehat{Q}_2(u)$, which we will specify later. After the Mellin transform
and again cancelling the overall coefficient $K_1 \, \sqrt{z(1-z)}$, we find for the $z$-space
counterpart of the above inhomogeneity
\be
\frac{1}{z(1-z)}  \, \widehat{\Psi}_2(z) \, .
\label{MB2}
\ee
Here $\widehat{Q}_2(u)$ and $\widehat{\Psi}_2(z)$ have the following expansions
\be
\widehat{Q}_2 (u) \, = \,
\sum_{k=1}^{M} \Re{\rm e}[ P_{k}(u)] \,
\, \widehat{F}_2(k)
\, , \qquad
\widehat{\Psi}_2 (z)   \, = \,
\frac{1}{2} \, \sum_{k=1}^{M} \, \Bigl[ z^k + (1-z)^k \Bigr] \,
\, \widehat{F}_2(k)
\, .
\label{MB3}
\ee
As it was discussed in the previous section, the expansions (\ref{MB3}) prevent the appearance
of negative powers of $z$ and/or $(1-z)$ in the right-hand side of the Baxter equation. Then the
corresponding particular solutions $\overline{Q}_2(u)$ and 
$\overline{\Psi}_2 (z)$ to the Baxter equation can be found
in a similar fashion
\be
\overline{Q}_2 (u) \, = \,
\sum_{k=0}^{M} \Re{\rm e}[ P_{k}(u)] \,
\, F_2(k)
\, , \qquad
\overline{\Psi}_2 (z)   \, = \,
\frac{1}{2} \, \sum_{k=0}^{M} \, \Bigl[ z^k + (1-z)^k \Bigr] \,
\, F_2(k) \, ,
\label{MB4}
\ee
where the coefficients $F_2(k)$ are determined by $\widehat{F}_2(k)$.

In the following, it is convenient to separately consider the cases of symmetric and
non-symmetric functions $\widehat{Q}_2(u)$ and $\widehat{\Psi}_2(z)$ with respect to
the replacements $u \to -u$ and $z \to (1-z)$, respectively.

\vspace{0.5cm}

\noindent {\sl Symmetric case.} In this case, 
$\widehat{\Psi}_2(1-z) = \widehat{\Psi}_2(z)$ and Eq.\ (\ref{MB2})
is equal to
\be
\frac{1}{z}  \, \widehat{\Psi}_2(z) \,
+ \, \frac{1}{(1-z)} \, \widehat{\Psi}_2(z)
\, .
\label{MB5}
\ee
Considering only the $z^k$-part of the expansion, we find from (\ref{MB5})
\be
\sum_{k=0}^{M-1} \, z^k \, \widehat{F}_2(k+1)
\, .
\label{MB6}
\ee
Then, introducing new coefficients via
\be
F_2(k) = R_{k}(M)f_2(k)
\, , \qquad
\widehat{F}_2(k) = R_{k}(M) \widehat{f}_2(k)\,,
\label{MB7}
\ee
we deduce from Eqs.\ (\ref{M26a}), (\ref{M27}) and (\ref{MB6}) the following representation
\be
f_2(k)= f_2(0) + 2 \sum_{m=1}^k \, \frac{\widehat{f}_2(m)}{m^2}
\, .
\label{MB7a}
\ee

If $\widehat{Q}_2(u) = Q_0$, the coefficients read
\be
\widehat{f}_2(k)= 1 \, , \qquad
f_2(k)= 2 S_2(k)
\, ,
\label{MB8}
\ee
and, thus, they reproduce the $Q_1$ Baxter function, when $\gamma_0 = \tilde{b}(M)=0$ (see (\ref{NN2+})).

\vspace{0.5cm}

\noindent {\sl Non-symmetric case.} As in the symmetric case, we can cast it in the form of the $z^k$-expansion,
\be
\frac{1}{z(1-z)}  \,
\sum_{k=1}^{M} \, z^k \, \widehat{F}_2(k) \, = \, \sum_{m=0}^{\infty} \, z^{m-1} \,
\sum_{k=1}^{M} \, z^k \, \widehat{F}_2(k) \, = \, \sum_{k=1}^{M} \, z^k \,
\widehat{F}_2(k) \, \sum_{p=k}^{\infty} \, z^{p-1} \, ,
\label{MB9}
\ee
where $p=m+k$, however, now being an infinite series rather than a finite sum. Splitting the series
$\sum_{p=k}^{\infty}$ into two parts
\be
\sum_{p=k}^{\infty} ~=~ \sum_{p=k}^{M} + \sum_{p=M+1}^{\infty} \, ,
\label{MB10}
\ee
we replace the right-hand side of (\ref{MB9}) by
\be
\sum_{p=1}^{M} \, z^{p-1} \, \widehat{\Phi}_2(p)
+ \sum^{\infty}_{p=M+1} \, z^{p-1} \, \widehat{\Phi}_2(M) \, ,
\label{MB11}
\ee
where
\be
\widehat{\Phi}_2(p) ~=~ \sum_{k=1}^{p} \, \widehat{F}_2(k) \, .
\label{MB12}
\ee
Then, Eq.\ (\ref{MB11}) can be represented as
\be
\sum_{p=1}^{M} \, z^{p-1} \, \Bigl[ \widehat{\Phi}_2(p) - \widehat{\Phi}_2(M) \Bigr]
+  \widehat{\Phi}_2(M) \, \sum^{\infty}_{p=1} \, z^{p-1}
\, = \, \sum_{p=1}^{M} \, z^{p-1} \, \Bigl[ \widehat{\Phi}_2(p) -
\widehat{\Phi}_2(M) \Bigr] +
\frac{\widehat{\Phi}_2(M)}{1-z}
\, .
\label{MB13}
\ee
As a consequence, Eq.\ (\ref{MB2}) reads
\be
\frac{1}{z(1-z)}  \, \widehat{\Psi}_2(z) ~=~ \frac{1}{2} \,
\sum_{p=1}^{M} \, \Bigl(z^{p-1} + (1-z)^{p-1} \Bigr) \,
\Bigl[ \widehat{\Phi}_2(p) -
\widehat{\Phi}_2(M) \Bigr] +
\frac{\widehat{\Phi}_2(M) }{2z(1-z)} \, .
\label{MB14}
\ee

To proceed further we have to address the nonpolynomial term in Eq.\ (\ref{MB14}). Notice that if we
add to the left-hand side of (\ref{M19}) the Baxter equation obeyed by the solution $Q_1(u)$
with the formal condition $\gamma_0 = 0$, this will yield the following contribution to the
inhomogeneity
\be
\frac{K_2}{z(1-z)}  \, \Psi_0(z)\, ,
\label{MB15}
\ee
where $\Psi_0(z)$ is the leading order solution of the Baxter equation with the property
\be
\Psi_0(z) ~=~ 1 + \widetilde{\Psi}_0(z)
, \qquad
\widetilde{\Psi}_0(z) ~=~ \sum_{k=1}^{M} \, z^k \, R_k(M)
\, .
\ee
Put together, we now have the right-hand side in the form
\be
\frac{1}{2} \,
\sum_{p=1}^{M} \, \Bigl(z^{p-1} + (1-z)^{p-1} \Bigr) \,
\Bigl[ \widehat{\Phi}_2(p) -
\widehat{\Phi}_2(M)  +2K_2 \, R_k(M)
\Bigr] +
\Bigl[\widehat{\Phi}_2(M)  +2K_2 \Bigr]\, \frac{1}{2z(1-z)} \, .
\label{MB16}
\ee
Choosing $K_2=-\widehat{\Phi}_2(M)/2$, we cancel the unwanted term $\sim 1/(z(1-z))$
and find that the resulting Baxter equation is obeyed by the function
\be
\overline{Q}_2 (u) - 
\tfrac{1}{2} \, \widehat{\Phi}_2(M) \, Q_1 (u)|_{\gamma_0=0} \,,
\label{MB17}
\ee
which admits the following Mellin-space transform
\be
\frac{1}{2} \,
\sum_{p=1}^{M} \, \Bigl(z^{p-1} + (1-z)^{p-1} \Bigr) \,
\Bigl[ \widehat{\Phi}_2(p) -
\widehat{\Phi}_2(M)\Bigl(1+  R_p(M) \Bigr)
\Bigr]
\, .
\label{MB18}
\ee
Next, using Eqs.\ (\ref{M26a}), (\ref{M27}) and (\ref{MB18}) we find the coefficients $f_2(k)$ to be
\be
f_2(k)= f_2(0) + \sum_{m=1}^k \, \frac{B_m}{m^2R_m(M)}
\, ,
\label{MB17new}
\ee
where
\be
B_m = \widehat{\Phi}_2(m) - \widehat{\Phi}_2(M) \Bigl(1+ R_m(M)\Bigr)
\, .
\label{MB18new}
\ee
Finally, taking the solution for $Q_1(u)$ from Eq.\ (\ref{NN2+}) with $\gamma_0=0$,
we find the solution for $Q_2 (u)$ in the form of Eq.\ (\ref{MB17new}) with
\be
B_m = \widehat{\Phi}_2(m) - \widehat{\Phi}_2(M)  - \widetilde{b}(M) \, R_m(M)
\, .
\label{MB18newer}
\ee

%%%%%%%%%%%%%%%%%%%%%%%%%%%%%%%%%%%%%%%%%%%%%%%%%%%%%%%%%%%%%%%%%%%%%%%
\subsubsection{Another form for $Q_1 (u)$}
%%%%%%%%%%%%%%%%%%%%%%%%%%%%%%%%%%%%%%%%%%%%%%%%%%%%%%%%%%%%%%%%%%%%%%%

As we explained earlier, the solution to the Baxter equation in the Mellin space is
a sum of $z^k$- and $(1-z)^k$-expansions (see Eq. (\ref{M16c})). However, the application
of this form as the input to inhomogeneities on the right-hand side of the Baxter equation is
not very convenient. It is a better choice to cast it solely in terms of the $z^k$-expansion,
\be
\Psi_{\ell} (z)   \, = \,
\sum_{k=0}^{M} \, z^k \,
\widetilde{R}_{\ell,k}(M) \, .
\label{MC2}
\ee

For the four-loop calculation, considered in this study, the non-polynomial inhomogeneities
contain only the functions $\Psi_{0} (z)$ and $\Psi_{1} (z)$. Moreover, the function 
$\Psi_{0} (z)$ 
and the $\widetilde{b}(M)$ as well as the $\gamma_0$ part of $\Psi_{1} (z)$ are symmetric under the
interchange $z \leftrightarrow (1-z)$ and do not represent a difficulty on their own.
Thus it is necessary to address the issue of non-symmetric contributions only for the remainder
of the function $\Psi_{1} (z)$, with the coefficient $\sim R_k(M) \, S_2(k)$.

Our goal is to find a relation of the form
\be
\frac{1}{2} \, \sum_{k=0}^{M} \, \Bigl[ z^k + (1-z)^k \Bigr] \,
\, R_k(M) \, S_{2}(k) = \sum_{k=0}^{M} \, z^k \,
\widetilde{R}_{2,k}(M) +  K_3 \Psi_0(z)
\, ,
\label{MC3}
\ee
where we also added the solution of the homogeneous Baxter equation with a coefficient $K_3$.
To start with, we note that the left-hand side of Eq.\ (\ref{MC3}) is a solution of the
Baxter equation with the following right-hand side
\be
\frac{1}{2} \, \sum_{k=1}^{M} \, \Bigl[ z^{k-1} + (1-z)^{k-1} \Bigr] \,
\, R_{k}(M)  \, .
\label{MC4}
\ee
The second term here, i.e.~$\sum_{k=0}^{M} \, (1-z)^{k-1} \, R_{k}(M)$, leads to the solution
of the Baxter equation in the form $\sum_{k=0}^{M} \, (1-z)^k \, R_{k}(M)\, S_{2}(k) $. 
Thus, we merely
have to replace the $(1-z)^k$-expansion by the $z^k$-expansion. Indeed, one can show,
that
\be
\sum_{k=1}^{M} \, (1-z)^{k-1} \, R_{k}(M) ~=~ \sum_{k=1}^{M} \, z^{k-1} \,
\sum_{m=1}^{k-1} R_{m}(M)\,,
\label{MC5}
\ee
and Eq.\ (\ref{MC4}) can be represented as
\be
\sum_{k=1}^{M} \, z^{k-1} \, \sum_{m=1}^{k} R_{m}(M) \, .
\label{MC6}
\ee
Thus, the general solution to Eq.\ (\ref{MC4}) admits the form
\be
\frac{1}{2} \, \sum_{k=0}^{M} \, \Bigl[ z^k + (1-z)^k \Bigr] \,
\, R_k(M) \, S_{2}(k) ~=~ \sum_{k=0}^{M} \, z^k \, R_k(M) \,
\, V_{2,0}(k)
+ K_3 \Psi_0(z)
\, ,
\label{MC7}
\ee
where $V_{2,0}(k)$ is given by Eq.\ (\ref{N4b10}) in the main text and the coefficient $K_3$
is fixed from numerical agreement between the left- and right-hand sides of Eq.\ (\ref{MC8})
\be
K_3 = -2S_{-2}(M)
\, .
\label{MC9}
\ee
The corresponding $u$-space form is then given by
\be
\sum_{k=0}^{M} \Re{\rm e}[ P_{k}(u)] \,
\, R_k(M) \, S_{2}(k) = \sum_{k=0}^{M} P_{k}(u) \,
 R_k(M) \, V_{2,0}(k)
+  K_3 \, Q_0(u) \, .
\label{MC8}
\ee

%%%%%%%%%%%%%%%%%%%%%%%%%%%%%%%%%%%%%%%%%%%%%%%%%%%%%%%%%%%%%%%%%%%%%%%
\subsection{Non-polynomial inhomogeneities}
%%%%%%%%%%%%%%%%%%%%%%%%%%%%%%%%%%%%%%%%%%%%%%%%%%%%%%%%%%%%%%%%%%%%%%%

Let us now address non-polynomial inhomogeneities emerging in the Baxter
equation. To start
with, we split the Baxter function (and its Mellin transform) at $\ell$-th order
as
\be
Q_\ell (u+ai) \, = \, R_{\ell, 0}(M) + \widetilde{Q}_\ell (u+ai)
\, , \qquad
\Psi_\ell (z) \, = \, R_{\ell, 0}(M) + \widetilde{\Psi}_\ell (z)
\, ,
\label{M22}
\ee
with $R_{0,0}(M) = 1$, where
\ba
\widetilde{Q}(u) \, = \,
\sum_{k=1}^{M} \, \widehat{R}_k(M) \, P_k(u)
\, , \qquad
\widetilde{\Psi}(z)   \, = \, K_1 \, \sum_{k=1}^{M} \, \widehat{R}_k(M) \, z^k,
\label{M22a}
\ea
Then the inverse Mellin transform of a generic inhomogeneity is given by
\ba
\label{M23}
\frac{L!}{(u^+)^{L+1}} \, \widetilde{Q}_\ell (u + i)
&+&
(-1)^{L+1} \frac{L!}{(u^-)^{L+1}} \, \widetilde{Q}_\ell (u - i)
\\
& \stackrel{M^{-1}}{\to}&
(-i)^{L+1} K \, \int_0^{\omega}  \frac{d\omega_1}{\omega_1} \,
{\left(\ln \frac{\omega}{\omega_1}\right)}^{L}  \,
{\left(\frac{\omega}{\omega_1}\right)}^{-b}  \,\biggl\{ -{\omega_1}^{\mp 1}
 \widetilde{Q}_\ell (-\omega_1) \biggr\}
\nonumber \\
&-&
i^{L+1} K \, \int_{\omega}^{\infty}  \frac{d\omega_1}{\omega_1}
\,
{\left(\ln \frac{\omega}{\omega_1}\right)}^{L}  \,
{\left(\frac{\omega}{\omega_1}\right)}^{-b}  \, \biggl\{ -{\omega_1}^{\mp 1}
\widetilde{Q}_\ell (-\omega_1) \biggr\}
\, . \nonumber
\ea
After the substitution (\ref{M13}) for $\omega \to z$, we have for the r.h.s. of
Eq.\
(\ref{M23})
\ba
- (-i)^{L+1} K_1 \, \sqrt{z(1-z)} \, \biggl[
&&
\frac{1}{z}\int_0^{z}  \frac{d z_1}{z_1} \,
{\left(\ln \frac{z(1-z_1)}{z_1(1-z)}\right)}^{L}  \,
\tilde{\Psi}_s(z_1)
\nonumber\\
+
&&
\frac{1}{(1-z)}\int_z^{1}  \frac{d z_1}{(1-z_1)} \,
{\left(\ln \frac{z_1(1-z)}{z(1-z_1)}\right)}^{L}  \,
\tilde{\Psi}_s(z_1) \biggr] \,.
\label{M24last}
\ea
The Baxter function $Q(u)$ and its Mellin transform $\Psi (z)$ possess symmetry
properties
preserved to all order of perturbation theory:
\be
\label{SymmProp}
Q (-u) = Q (u)
\, , \qquad
\Psi (1-z) = \Psi (z)
\, .
\ee
Thus, in all calculations we will focus on just one term in Eq.\ (\ref{M24last}),
say, the first
one on the right-hand side, and then get the complete result for the Baxter
function obeying
the required symmetry properties (\ref{SymmProp}) by adding contribution with
reflected
argument. With this argument in mind, Eqs.\ (\ref{M23}) and (\ref{M24last}) can be
rewritten as
\ba
\label{M26}
\frac{L!}{(u^+)^{L+1}} \, \widetilde{Q}_\ell (u+ i)
&+&
(-1)^{L+1} \frac{L!}{(u^-)^{L+1}} \, \widetilde{Q}_\ell (u- i)
\,   \stackrel{M^{-1}}{\to} \,
\\
&-&
2 (-i)^{L+1} K_1 \, \frac{[z(1-z)]^{1/2}}{z}
\int_0^{z}  \frac{d z_1}{z_1} \,
{\left(\ln \frac{z(1-z_1)}{z_1(1-z)}\right)}^{L}  \,
\widetilde{\Psi}_\ell (z_1)
\, .
\nonumber
\ea
To calculate the integral in the right-hand side of (\ref{M26}), it is
convenient to re-express
the integrand involving the logarithm as a rational function of its argument via
\be
{\left(\ln \frac{z(1-z_1)}{z_1(1-z)}\right)}^{L}  \, = \,
{\left(\frac{d}{d \varepsilon}\right)}^L \,
{\left( \frac{z(1-z_1)}{z_1(1-z)}\right)}^\varepsilon \biggl|_{\varepsilon = 0}
\, .
\label{M28}
\ee
Now, using the expansion for $\widetilde{\Psi}_\ell (z)$, after some algebra we
obtain
\be
\int_0^{z}  \frac{d z_1}{z_1} \,
{\left( \frac{z(1-z_1)}{z_1(1-z)}\right)}^\varepsilon \,
\widetilde{\Psi}_\ell (z_1)
\, = \, \sum_{p=1}^{\infty} \, z^p \, \biggl[ \frac{R_{p,\ell}(M)}{p -
\varepsilon}
+
\frac{\varepsilon \Gamma(p)}{\Gamma(p+1-\varepsilon)} \, \sum_{k=1}^{p-1} \,
R_{k,\ell} (M) \,
\frac{\Gamma(k-\varepsilon)}{k!} \biggr]
\, .
\label{M30}
\ee
The first term on the right-hand side of this equation is obviously polynomial,
but the second
one is not. Note, however, that for $p>M$ we can represent it as
\ba
\label{M31}
\sum_{p=M+1}^{\infty} \, z^p \,
\frac{\varepsilon \Gamma(p)}{\Gamma(p+1-\varepsilon)} \,
&& \sum_{k=1}^{p-1} \, R_{k,\ell}(M) \,
\frac{\Gamma(k-\varepsilon)}{k!}  \\
= &&
\sum_{p=M+1}^{\infty} \, z^p \,
\frac{\varepsilon \Gamma(p)}{\Gamma(p+1-\varepsilon)} \, \sum_{k=1}^{M} \,
R_{k,\ell}(M) \,
\frac{\Gamma(k-\varepsilon)}{k!} \, , \nonumber
\ea
where the inner sum in the last term can be cast in the form
\be
\sum_{k=1}^{M} \, R_{k,\ell}(M) \,
\frac{\Gamma(k-\varepsilon)}{k!\Gamma(-\varepsilon)} \, = \,
\widetilde{Q}_\ell\left(u=\ft{i}{2}+i\varepsilon\right) \, = \,
Q_\ell\left(u=\ft{i}{2}+i\varepsilon\right) - R_{0,s}(M)
%1
\, .
\label{M32}
\ee
Thus, we have
\ba
\int_0^{z}  \frac{d z_1}{z_1} \,
{\left( \frac{z(1-z_1)}{z_1(1-z)}\right)}^\varepsilon \,
\tilde{\Psi}_\ell(z_1)
&=& \sum_{p=1}^{M} \, z^p \, \biggl[ \frac{R_{p,\ell}(M)}{p-\varepsilon}
+
\frac{\varepsilon \Gamma(p)}{\Gamma(p+1-\varepsilon)} \, \sum_{k=1}^{p-1} \,
R_{k,\ell}(M) \,
\frac{\Gamma(k-\varepsilon)}{k!} \biggr] \nonumber \\
&-&  \,
\sum_{p=M+1}^{\infty} \, z^p \,
\frac{\Gamma(p)\Gamma(1-\varepsilon)}{\Gamma(p+1-\varepsilon)} \,
\widetilde{Q}_\ell \left(u=\ft{i}{2}+i \varepsilon\right)
\, .
\label{M32_2}
\ea
To get back to the integrand in question, we have to differentiate both sides
w.r.t.\ $\varepsilon$
and set it to zero. Then one immediately finds that the last term becomes
\be
-\sum_{k=0}^L \, C^k_L \,
\sum_{p=M+1}^{\infty} \, \frac{z^p}{p} \, i^{k} 
\widetilde{Q}_\ell^{(k)}(\ft{i}{2})
\,
{\left(\frac{d}{d\varepsilon}\right)}^{L-k} \,
\left\{\frac{\Gamma(p)\Gamma(1-\varepsilon)}{\Gamma(p+1-
\varepsilon)}\right\}\biggl|_{\varepsilon=0},
\label{M33}
\ee
where we 
%used the fact that $\widetilde{Q}_\ell (u=\ft{i}{2}) = 0$ and
introduced notations for
\be
C^k_L \, = \, \frac{L!}{k!(L-k)!}
\, , ~~
%\qquad
\widetilde{Q}_{\ell}^{(k)}(\pm \ft{i}{2}) \, = \,
{\left(\frac{d}{du}\right)}^k \, 
\widetilde{Q}_\ell \left(u=\pm \ft{i}{2}\right)
%\, , 
\, , ~~
%\qquad
\widetilde{Q}_{\ell}^{(0)}(\pm \ft{i}{2}) \, = \,
\widetilde{Q}_{\ell}(\pm \ft{i}{2}) \, ,  
\label{M34}
\ee
and for the coefficients
\be
Z_L(p,k) \, = \,
{\left(\frac{d}{d \varepsilon}\right)}^{L} \,
\left\{\frac{\Gamma(k+1-\varepsilon)}{\Gamma(p+1-
\varepsilon)}\right\}\biggl|_{\varepsilon=0}
\, ,
\label{M35}
\ee
which are expressed in terms of harmonic numbers as follows for a few low values
of $L$,
\ba
Z_0(p,k) &=& 1 \, , \nonumber\\
Z_1(p,k) &=& S_1(p,k)
\, , \nonumber\\
Z_2(p,k) &=& S^2_1(p,k) + S_2(p,k), \nonumber \\
Z_3(p,k) &=& S^3_1(p,k) + 3 S_1(p,k)S_2(p,k) + 2 S_3(p,k)
\, , \nonumber \\
Z_4(p,k) &=& S^4_1(p,k) + 6 S^2_1(p,k)S_2(p,k) + 8 S_1(p,k)S_3(p,k)
+3S^2_2(p,k)+ 6 S_4(p,k)
\, , \nonumber \\
Z_5(p,k) &=& S^5_1(p,k) + 10 S^3_1(p,k)S_2(p,k) + 20 S^2_1(p,k)S_3(p,k)
+15S_1(p,k)S^2_2(p,k)
\nonumber \\
&+&
30 S_1(p,k)S_4(p,k) + 20 S_2(p,k)S_3(p,k) + 24 S_5(p,k)
\, ,
\label{M36}
\ea
where
\be
S_L(p,k) \, = \, S_L(p) - S_L(k)
\ee
and $Z_L(p) \, = \, Z_L(p,0)$.

Assembling all results, we have
\ba
&& \frac{1}{(u^+)^{L+1}} \, \widetilde{Q}_\ell (u+ i)
+
\frac{(-1)^{L+1}}{(u^-)^{L+1}} \, \widetilde{Q}_\ell (u- i)
\,   \stackrel{M^{-1}}{\to} \, \nonumber \\
&&\quad -2 (-i)^{L+1} \, K_1 \sqrt{z(1-z)} \, \biggl[
\sum_{p=1}^{M} \, \frac{z^{p-1}}{p}  \biggl\{ \frac{R_{p,\ell}(M)}{p^L}
 +
\frac{1}{(L-1)!} \, \sum_{k=1}^{p-1} \, \frac{R_{k,\ell}(M)}{k} \, Z_{L-1}(p,k-
1)
\biggr\}  \nonumber \\
&&\qquad\qquad\qquad\qquad\qquad\quad -  \,
\sum_{k=0}^L \,\frac{i^l}{k!(L-k)!} \,  \widetilde{Q}_{\ell}^{(k)}(\ft{i}{2})
\,
\sum_{p=M+1}^{\infty} \, \frac{z^{p-1}}{p} \,  Z_{L-k}(p) \biggr]
\, .
\label{M37}
\ea
And after returning to the $u$-space the final result, free from Stirling
numbers, reads
\ba
\frac{1}{(u^+)^{L+1}} \, \widetilde{Q}_\ell (u+ i)
&+&
\frac{(-1)^{L+1}}{(u^-)^{L+1}} \, \widetilde{Q}_\ell (u- i)
\nonumber \\
= -2 (-i)^{L+1} \, \biggl[
&&\!\!\!
\sum_{p=1}^{M} \, \frac{P_{p-1}(u)}{p} \, \biggl\{ \frac{R_{p,\ell}(M)}{p^L}
\, +
\frac{1}{(L-1)!} \, \sum_{k=1}^{p-1} \, \frac{R_{k,\ell}(M)}{k} \, Z_{L-1}(p,k-
1)
\biggr\}  \nonumber \\
-
&&\!\!\!
\sum_{k=0}^L \,\frac{i^k}{k!(L-k)!} \,   \widetilde{Q}_{\ell}^{(k)}(\ft{i}{2})
\,
\sum_{p=M+1}^{\infty} \, \frac{P_{p-1}(u)}{p} \,  Z_{L-k}(p) \biggr]
\, ,
\label{M38}
\ea
where $P_{p-1}(u)$ is given in Eq.  (\ref{N1}). Note that for $L>0$, Eq.\
(\ref{M38}) can be
simplified to
\ba
\frac{1}{(u^+)^{L+1}} \, \widetilde{Q}_\ell (u+ i)
&+&
\frac{(-1)^{L+1}}{(u^-)^{L+1}} \, \widetilde{Q}_\ell (u- i)
\nonumber \\
= -2 (-i)^{L+1} \, \biggl[
&&\!\!\! \frac{1}{(L-1)!} \,
\sum_{p=1}^{M} \, \frac{P_{p-1}(u)}{p} \,
\sum_{k=1}^{p} \, \frac{R_{k,\ell}(M)}{k} \, Z_{L-1}(p,k-1)  \,
\nonumber \\
-
&&\!\!\!
\sum_{k=0}^L \,\frac{i^k}{k!(L-k)!} \,   \widetilde{Q}_{\ell}^{(k)}(\ft{i}{2})
\,
\sum_{p=M+1}^{\infty} \, \frac{P_{p-1}(u)}{p} \,  Z_{L-k}(p) \biggr]
\, .
\label{M38a}
\ea

Repeating all above calculations, it is possible to obtain in an analogous fashion
\ba
\frac{1}{(u^+)^{L+1}} \, \widetilde{Q}_\ell (u)
&+&
\frac{(-1)^{L+1}}{(u^-)^{L+1}} \, \widetilde{Q}_\ell (u)
\label{M39} \\
&\stackrel{M^{-1}}{\to}&
2 (-i)^{L+1} \, K_1 \, \sqrt{z(1-z)} \, \biggl[\frac{1}{L!}
\sum_{p=1}^{M} \, \frac{z^{p-1}}{p} \,  \sum_{k=1}^{p-1} \,
\frac{R_{k,\ell}(M)}{k} \, Z_L(p,k)
\nonumber\\
&&\qquad
+
\,
\sum_{k=0}^L \,\frac{i^k }{k!(L-k)!}
\,   \widetilde{Q}_{\ell}^{(k)}(-\ft{i}{2})
\,
\sum_{p=M+1}^{\infty} \, \frac{z^{p-1}}{p} \,  Z_{L-k}(p) \biggr]
\, ,
\nonumber
\ea
and as a consequence
\ba
\label{M40}
\frac{1}{(u^+)^{L+1}} \, \widetilde{Q}_\ell (u)
&+&
\frac{(-1)^{L+1}}{(u^-)^{L+1}} \, \widetilde{Q}_\ell (u)
\\
&=&
2 (-i)^{L+1} \, \biggl[\frac{1}{L!}
\, \sum_{p=1}^{M} \, \frac{P_{p-1}(u)}{p} \,  \sum_{k=1}^{p-1} \,
%\frac{R_{k,\ell}(M)}{k} \, Z_L(p,k)
R_{k,\ell}(M) \, Z_L(p,k)
\nonumber \\
&&\qquad
+
\sum_{k=0}^L \,\frac{i^k}{k!(L-k)!} \,   \widetilde{Q}_{\ell}^{(k)}(-\ft{i}{2})
\,
\sum_{p=M+1}^{\infty} \, \frac{P_{p-1}(u)}{p} \,  Z_{L-k}(p) \biggr]
\, . \nonumber
\ea
Last but not least, as a particular case of (\ref{M39}) and (\ref{M40}), we find
\ba
&& \frac{1}{(u^+)^{L+1}} \, + \frac{(-1)^{L+1}}{(u^-)^{L+1}}
\, \stackrel{M^{-1}}{\to}
\, 2 (-i)^{L+1} \, K_1 \, \sqrt{z(1-z)} \,
\sum_{p=1}^{\infty} \, \frac{z^{p-1}}{p} \,  Z_{L}(p)
\, , \label{M41}\\
&& \frac{1}{(u^+)^{L+1}} \, + \frac{(-1)^{L+1}}{(u^-)^{L+1}}
\,  = \, 2 (-i)^{L+1} \,
\sum_{p=1}^{\infty} \, \frac{P_{p-1}(u)}{p} \,  Z_{L}(p)
\, . \label{M42}
\ea
%%%%%%%%%%%%%%%%%%%%%%%%%%%%%%%%%%%%%%%%%%%%%%%%%%%%%%%%%%%%%%%%%%%%%%%
\section{On the degree reducing constants}
%%%%%%%%%%%%%%%%%%%%%%%%%%%%%%%%%%%%%%%%%%%%%%%%%%%%%%%%%%%%%%%%%%%%%%%
We have seen that the polynomial-type contribution to the Baxter functions takes
the form of a linear combination
of higher derivatives $T_{(a, b, c, \dots)}$ plus a constant times the leading
order Baxter polynomial $Q_{0}$.
This constant is fixed by the requirement that the monomial $u^M$ cancels in the
total contribution.
Thus, the following ratios are all we need to determine the degree reducing
constant
\be
\rho_{(a, b, c, \dots)} = \frac{\mbox{coefficient of $u^M$ in $T_{(a, b, c,
\dots)}$}}
{\mbox{coefficient of $u^M$ in $Q_{0}$}}.
\ee
Here, we list the cases which are needed for computing $a_2$, and $a_3$. The
notation is
\be
S_{a, b, \dots} = S_{a, b, \dots}(M),\quad
\widetilde S_{a, b, \dots} = S_{a, b, \dots}(2M),\quad
\widehat S_{a, b, \dots} = S_{a, b, \dots}(M/2).
\ee
\baa
\rho_{(1,0)} &=& 2 \widetilde{S}_1-3 S_1, \\
\rho_{(0,1)} &=& 2 \widehat{S}_2, \\
\rho_{(2,0)} &=& 9 S_1^2-12 \widetilde{S}_1 S_1+4 \widetilde{S}_1^2+5 S_2-4
\widetilde{S}_2, \\
\rho_{(1,1)} &=& -6 S_1 \widehat{S}_2+4 \widetilde{S}_1 \widehat{S}_2-2
\widehat{S}_3, \\
\rho_{(0,2)} &=& 12 \widehat{S}_2^2+12 \widehat{S}_4, \\
\rho_{(3,0)} &=& -27 S_1^3+54 \widetilde{S}_1 S_1^2-36 \widetilde{S}_1^2 S_1-45
S_2 S_1+36 \widetilde{S}_2 S_1+\nonumber\\
&& + 8 \widetilde{S}_1^3-18 S_3+30 S_2 \widetilde{S}_1-24 \widetilde{S}_1
\widetilde{S}_2+16 \widetilde{S}_3, \\
\rho_{(2,1)} &=& 18 \widehat{S}_2 S_1^2+12 \widehat{S}_3 S_1-24 \widehat{S}_2
\widetilde{S}_1 S_1+2 \widehat{S}_2^2+8 \widehat{S}_2
\widetilde{S}_1^2+\nonumber\\
&& + 10 S_2 \widehat{S}_2+6 \widehat{S}_4-8 \widehat{S}_3 \widetilde{S}_1-8
\widehat{S}_2 \widetilde{S}_2, \\
\rho_{(1,2)} &=& -36 S_1 \widehat{S}_2^2+24 \widetilde{S}_1 \widehat{S}_2^2-24
\widehat{S}_3 \widehat{S}_2-36 S_1 \widehat{S}_4-24 \widehat{S}_5+24
\widehat{S}_4 \widetilde{S}_1, \\
\rho_{(0,3)} &=& 120 \widehat{S}_2^3+360 \widehat{S}_4 \widehat{S}_2+240
\widehat{S}_6.
\eaa
In the case of $b_3$, we need the ratios
\baa
\rho_{(0,0,2)} &=& 2 S_2, \\
\rho_{(0,0,4)} &=& 12 S_2^2+12 S_4, \\
\rho_{(0,0,4)} &=& 12 S_2^2+12 S_4, \\
\rho_{(0,0,4)} &=& 12 S_2^2+12 S_4, \\
\rho_{(0,0,6)} &=& 120 S_2^3+360 S_4 S_2+240 S_6, \\
\rho_{(0,2,0)} &=& 4 S_1^2+2 S_2, \\
\rho_{(0,2,2)} &=& 8 S_2 S_1^2+16 S_3 S_1+4 S_2^2+12 S_4, \\
\rho_{(0,2,2)} &=& 8 S_2 S_1^2+16 S_3 S_1+4 S_2^2+12 S_4, \\
\rho_{(0,2,4)} &=& 24 S_2^3+48 S_1^2 S_2^2+192 S_1 S_3 S_2+168 S_4 S_2+96
S_3^2+48 S_1^2 S_4+\nonumber\\
&& + 192 S_1 S_5+240 S_6, \\
\rho_{(0,3,0)} &=& -8 S_1^3-12 S_2 S_1-4 S_3, \\
\rho_{(0,3,2)} &=& -16 S_2 S_1^3-48 S_3 S_1^2-24 S_2^2 S_1-72 S_4 S_1-32 S_2
S_3-48 S_5, \\
\rho_{(0,4,0)} &=& 16 S_1^4+48 S_2 S_1^2+32 S_3 S_1+12 S_2^2+12 S_4, \\
\rho_{(0,4,0)} &=& 16 S_1^4+48 S_2 S_1^2+32 S_3 S_1+12 S_2^2+12 S_4, \\
\rho_{(0,4,0)} &=& 16 S_1^4+48 S_2 S_1^2+32 S_3 S_1+12 S_2^2+12 S_4, \\
\rho_{(0,4,2)} &=& 32 S_2 S_1^4+128 S_3 S_1^3+96 S_2^2 S_1^2+288 S_4 S_1^2+256
S_2 S_3 S_1+\nonumber\\
&& + 384 S_5 S_1+24 S_2^3+64 S_3^2+168 S_2 S_4+240 S_6, \\
\rho_{(0,5,0)} &=& -32 S_1^5-160 S_2 S_1^3-160 S_3 S_1^2-120 S_2^2 S_1-120 S_4
S_1-80 S_2 S_3-48 S_5, \\
\rho_{(0,6,0)} &=& 64 S_1^6+480 S_2 S_1^4+640 S_3 S_1^3+720 S_2^2 S_1^2+720 S_4
S_1^2+960 S_2 S_3 S_1+\nonumber\\
&& + 576 S_5 S_1+120 S_2^3+160 S_3^2+360 S_2 S_4+240 S_6, \\
\rho_{(1,0,0)} &=& \tilde{S}_1-2 S_1, \\
\rho_{(1,0,2)} &=& -4 S_1 S_2+2 \tilde{S}_1 S_2-2 S_3, \\
\rho_{(1,0,4)} &=& -24 S_1 S_2^2+12 \tilde{S}_1 S_2^2-24 S_3 S_2-24 S_1 S_4-24
S_5+12 S_4 \tilde{S}_1, \\
\rho_{(1,2,0)} &=& -8 S_1^3+4 \tilde{S}_1 S_1^2-8 S_2 S_1-2 S_3+2 S_2
\tilde{S}_1, \\
\rho_{(1,2,2)} &=& -16 S_2 S_1^3-40 S_3 S_1^2+8 S_2 \tilde{S}_1 S_1^2-16 S_2^2
S_1-48 S_4 S_1+16 S_3 \tilde{S}_1 S_1+\nonumber\\
&& -16 S_2 S_3-24 S_5+4 S_2^2 \tilde{S}_1+12 S_4 \tilde{S}_1, \\
\rho_{(1,3,0)} &=& 16 S_1^4-8 \tilde{S}_1 S_1^3+36 S_2 S_1^2+20 S_3 S_1-12 S_2
\tilde{S}_1 S_1+6 S_2^2+6 S_4-4 S_3 \tilde{S}_1, \\
\rho_{(1,4,0)} &=& -32 S_1^5+16 \tilde{S}_1 S_1^4-128 S_2 S_1^3-112 S_3 S_1^2+48
S_2 \tilde{S}_1 S_1^2-72 S_2^2 S_1 + \nonumber\\
&& -72 S_4 S_1+32 S_3 \tilde{S}_1 S_1-40 S_2 S_3-24 S_5+12 S_2^2 \tilde{S}_1+12
S_4 \tilde{S}_1, \\
\rho_{(2,0,0)} &=& 4 S_1^2-4 \tilde{S}_1 S_1+\tilde{S}_1^2+2 S_2-\tilde{S}_2, \\
\rho_{(2,0,2)} &=& 8 S_2 S_1^2+8 S_3 S_1-8 S_2 \tilde{S}_1 S_1+6 S_2^2+2 S_2
\tilde{S}_1^2+6 S_4-4 S_3 \tilde{S}_1-2 S_2 \tilde{S}_2, \\
\rho_{(2,2,0)} &=& 16 S_1^4-16 \tilde{S}_1 S_1^3+4 \tilde{S}_1^2 S_1^2+32 S_2
S_1^2-4 \tilde{S}_2 S_1^2+16 S_3 S_1+\nonumber\\
&& -16 S_2 \tilde{S}_1 S_1+6 S_2^2+2 S_2 \tilde{S}_1^2+6 S_4-4 S_3 \tilde{S}_1-2
S_2 \tilde{S}_2, \\
\rho_{(3,0,0)} &=& -8 S_1^3+12 \tilde{S}_1 S_1^2-6 \tilde{S}_1^2 S_1-12 S_2
S_1+6 \tilde{S}_2 S_1+\tilde{S}_1^3+\nonumber\\
&& -4 S_3+6 S_2 \tilde{S}_1-3 \tilde{S}_1 \tilde{S}_2+2 \tilde{S}_3.
\eaa

%%%%%%%%%%%%%%%%%%%%%%%%%%%%%%%%%%%%%%%%%%%%%%%%%%%%%%%%%%%%%%%

%%%%%%%%%%%%%%%%%%%%%%%%%%%%%%%%%%%%%%%%%%%%%%%%%%%%%%%%%%%%%%%%%%%%

\end{document}